\documentclass[aps,11pt,prd,notitlepage,tightenlines,nofootinbib,showpacs]{revtex4-1}
\usepackage{amsmath, amssymb, amsthm}
\usepackage{mathtools}
\usepackage{mathrsfs}
\usepackage{bm}
\usepackage{slashed}   
\usepackage{graphicx}
\usepackage{multirow}
\usepackage{tikz}
\usepackage[caption=false]{subfig}
\usepackage{relsize}	
\usepackage{array}
\usepackage{float}
\usepackage{color}
\usepackage{xcolor}
\usepackage{soul}
\usepackage{siunitx}
\usepackage{hyperref}
\hypersetup{colorlinks=true,linkcolor=red,anchorcolor=blue,citecolor=darkgreen, filecolor=blue,urlcolor=red,bookmarksnumbered=true,
pdfview=FitB
}

\def\dd{{\mathrm{d}}}
\def\imag{{\mathrm{i}}}

\mathchardef\-="2D

\newcommand{\half}[1][1] {\mathsmaller{\frac{#1}{2}}}

\makeatletter
\newcommand*{\transpose}{%
  {\mathpalette\@transpose{}}%
}
\newcommand*{\@transpose}[2]{%
  \raisebox{\depth}{$\m@th#1\intercal$}%
}
\makeatother

\colorlet{darkgreen}{green!60!black}
\colorlet{brightyellow}{yellow!75!red}
\colorlet{orange}{red!50!yellow}
\colorlet{darkblue}{blue!80!green}
\colorlet{darkred}{red!80!black}
\colorlet{greenblue}{green!50!blue}

\makeatletter

\newcommand{\Rmnum}[1]{\expandafter\@slowromancap\romannumeral #1@}
\makeatother

\begin{document}
\title{Quarkonium as relativistic bound state on the light front}

\author{Yang~Li}
\email{leeyoung@iastate.edu} 
\affiliation{Department of Physics and Astronomy, Iowa State University, Ames, IA 50011, USA}
%
\author{Pieter~Maris}
\affiliation{Department of Physics and Astronomy, Iowa State University, Ames, IA 50011, USA}
%
%
\author{James~P.~Vary}
\affiliation{Department of Physics and Astronomy, Iowa State University, Ames, IA 50011, USA}

\date{\today}

\begin{abstract}
We study charmonium and bottomonium as relativistic bound states in a light-front quantized Hamiltonian
formalism. The effective Hamiltonian is based on light-front holography. We use a recently proposed longitudinal
confinement to complete the soft-wall holographic potential for the heavy flavors. The spin structure
is generated from the one-gluon exchange interaction with a running coupling. The adoption of asymptotic 
freedom improves the spectroscopy compared with previous light-front 
results. 
Within this model, we compute the mass spectroscopy, decay constants and the r.m.s. radii. We also present 
a detailed study of the obtained light-front wave functions and use the wave functions to compute the light-cone
distributions, specifically the distribution amplitudes and parton distribution functions. 
Overall, our model provides a reasonable description of the heavy quarkonia. 
\end{abstract}

\maketitle 

\section{Introduction}

Non-perturbative calculations of quantum chromodynamics (QCD) provide insights into the fundamental structure of 
hadrons which constitute the majority of the visible matter in the Universe. 
Lattice gauge theory has produced high precision results for hadron spectroscopy and many other observables. It is expected that
Lattice QCD will eventually provide a valid description of the experimental data arising from both the theoretical progress and the growth
of computational capacity. 
On the other hand, QCD at high energy is most conveniently expressed through the light-front variables \cite{Lepage:1980fj}. While the
so-called ``hard processes'' may be evaluated through perturbation theory (pQCD), non-perturbative information from QCD is also needed and
is encoded within the so-called ``light-cone distributions''.
The light-cone distributions are intrinsically \emph{Minkowskian}, and cannot be easily extracted from a Euclidean formulation of quantum
field theories. It is anticipated that the light-front\footnote{In this article, we use the words ``light-front'' and ``light-cone''
interchangeably.} Hamiltonian formalism provides a complementary alternative to lattice gauge theory \cite{Bakker:2013cea}, with
convenient access to light-cone distributions and other observables. 

In principle, the hadron mass spectrum and light-front wave functions (LFWFs) can be obtained from diagonalizing the light-front quantized
QCD (LFQCD) Hamiltonian operator \cite{Brodsky:1997de}. Ab initio light-front Hamiltonian approaches, such as Discretized
Light-Cone Quantization (DLCQ, \cite{Pauli:1985ps}) and Basis Light-Front Quantization (BLFQ, \cite{Vary:2009gt}), have made important
strides in tackling various test problems, and show promise of advancing towards more realistic field theories, including QCD
\cite{Vary:2016emi}. As a complementary method to these ab initio approaches, light-front holography constructs an effective
Hamiltonian based on insights from string theory, and has been shown, notwithstanding criticisms (e.g., \cite{Ballon-Bayona:2014oma,
Glozman:2009bt}), to be a valuable approximation to QCD \cite{Brodsky:2014yha}.  The efforts to improve light-front holography can be
roughly cast into two categories: one is on the holographic QCD side (see \cite{Brodsky:2014yha} and the references therein); the other is
on the light-front Hamiltonian side (see \cite{Hiller:2016itl} for a recent review).

The present work falls into the second category. We generalize the light-front holographic QCD of Brodsky and de T\'eramond to incorporate
quark masses and quarkonium spin structure by extending the ``soft-wall'' light-front Hamiltonian. Our model introduces a phenomenological
effective Hamiltonian. Key elements include a confining potential in the longitudinal direction and an effective one-gluon exchange
interaction derived from light-front QCD \cite{Li:2015zda, Li:2016wwu}. 
It was long pointed out by Lepage and Brodsky \cite{Lepage:1980fj} that the dominant ultraviolet (UV) physics can be analyzed through
one-gluon exchange. Here, we combine the one-gluon exchange physics at short distance and the holographic QCD at long distance. 
The present work improves our previous calculation \cite{Li:2015zda} by including the evolution of the strong coupling as a function of
invariant 4-momentum transfer. Incorporating the
running coupling not only implements important QCD physics, but also improves the UV asymptotics
of the kernel. In particular, a previous non-covariant UV counterterm is now removed and the hyperfine structure is readily
improved as we present in this work.

The motivation of the present work is multi-fold. As stated, we supplement the light-front holographic QCD interaction with one-gluon
exchange, rather than patching the holographic wave functions with, e.g., spin structures (see, e.g., Ref.~{\cite{Chen:2016dlk}
and the references therein)}. The spectroscopy and the wave functions are obtained as a natural output. More importantly, we solve the
problem using the basis function method \cite{Vary:2009gt}. Effectively, we are applying BLFQ to a phenomenological interaction that
emulates features of QCD. Indeed, this work is a direct extension of the BLFQ approach to positronium in QED \cite{Wiecki:2014ola}. Finally,
we acknowledge the similarities between our work and the relativistic bound-state models in QCD (e.g., Refs.~\cite{Sommerer:1994bk,
Spence:1995bm, Spence:1999db, Maris:1999nt, Leitao:2016bqq}), especially the light-front QCD bound-state models \cite{Pauli:1996ne,
Brisudova:1995hv, Brisudova:1996vw, Glazek:2003ky, Glazek:2006cu, Choi:2015ywa}. 

We organize this paper as follows. In Sect.~\ref{sec:ingredients}, we introduce the theoretical model, including the longitudinal
confinement and a running strong coupling. The formulation and the methods are detailed in Sect.~\ref{sect:formalism}.
Sect.~\ref{sect:numerical_results} summarizes and analyzes the numerical results, including the spectroscopy, decay constants and
radii. Sect.~\ref{sect:lfwfs} presents LFWFs and light-cone distributions computed from them. We summarize the paper in
Sect.~\ref{sect:summary}.

\section{Holographic Confinement plus One-Gluon Exchange}\label{sec:ingredients}

We extend light-front holography by introducing realistic QCD interactions such as the one-gluon exchange interaction with running
coupling \cite{Li:2015zda}. In addition we include finite quark masses, important for heavy flavors, as well as a longitudinal confining
potential to complement the transverse holographic confining potential. Spin structure and excited states (radial and angular) naturally
emerge from the one-gluon exchange and its non-perturbative interplay with the confining potential \cite{Li:2016wwu}.
The effective Hamiltonian $H_\text{eff}\equiv P^+P^-_\text{eff}-\vec P^2_\perp$ reads, 
\begin{multline}\label{eqn:Heff}
H_\mathrm{eff} = \frac{\vec k^2_\perp + m_q^2}{x} + \frac{\vec k^2_\perp+m_{\bar q}^2}{1-x}
+ \kappa^4 \vec \zeta_\perp^2 - \frac{\kappa^4}{(m_q+m_{\bar q})^2} \partial_x\big( x(1-x) \partial_x \big)  \\
- \frac{C_F4\pi\alpha_s(Q^2)}{Q^2}\bar u_{s'}(k')\gamma_\mu u_s(k) \bar v_{\bar s}(\bar k) \gamma^\mu v_{\bar s'}(\bar k').
\end{multline}
where $\vec \zeta_\perp \equiv \sqrt{x(1-x)} \vec r_\perp$ is Brodsky and de T\'eramond's holographic variable \cite{Brodsky:2014yha},
$\partial_x f(x, \vec\zeta_\perp) = \partial f(x, \vec \zeta_\perp)/\partial x|_{\vec\zeta}$, $C_F = (N_c^2-1)/(2N_c)=4/3$ is the color
factor for the color singlet state. $\kappa$ is the strength of the confinement, and $m_q$ ($m_{\bar q}$) is the mass of the quark (anti-quark). 
$Q^2 = -(1/2) (k'-k)^2- (1/2) (\bar k'-\bar k)^2$ is the average 4-momentum squared carried by the exchanged
gluon. In terms of kinematical variables, 
\begin{equation}
 Q^2 = \frac{1}{2}\Big(\sqrt{\frac{x'}{x}}\vec k_\perp-\sqrt{\frac{x}{x'}}\vec k'_\perp\Big)^2 +
\frac{1}{2}\Big(\sqrt{\frac{1-x'}{1-x}}\vec k_\perp-\sqrt{\frac{1-x}{1-x'}}\vec k'_\perp \Big)^2 
+ \frac{1}{2}(x-x')^2\Big( \frac{m_q^2}{xx'}+\frac{m_{\bar q}^2}{(1-x)(1-x')}\Big) + \mu_g^2.
\end{equation}

\subsection{Longitudinal Confinement}

In Eq.~(\ref{eqn:Heff}), the term $\kappa^4 \vec\zeta^2_\perp \equiv \kappa^4 x(1-x)\vec r^2_\perp$ is the ``soft-wall'' confinement from
light-front holography, which is introduced in the massless case. For heavy quarkonium, the quark
masses and the longitudinal dynamics cannot be ignored\footnote{Without a longitudinal confinement, the longitudinal excitations will not be
separated by mass gaps. In light-front holography (no quark mass nor one-gluon exchange), these excitations are degenerate and the system is
two-dimensional in nature.} and we introduce a longitudinal confining interaction to complete the transverse holographic confinement. The
form of the longitudinal confinement is designed to produce a power-law behavior for the distribution amplitudes $\phi(x)\sim x^a(1-x)^b$ at
the endpoints (cf. \cite{Gutsche:2012ez, Gutsche:2013zia, Chabysheva:2012fe}). 

We fix the strength of the longitudinal confinement by matching to the transverse holographic confinement in the non-relativistic
limit. Therefore, rotational symmetry is retained in the heavy-quark limit.
Another advantage of this choice for the longitudinal confinement is that it produces, without the one-gluon exchange, analytic solutions.
Therefore, it affords computational convenience within the basis function method (see Sect.~\ref{sec:3.1}). 
In the massless limit, our wave function (without the one-gluon exchange) reduces to the soft-wall wave function of Brodsky and
de~T\'eramond\footnote{See Eq.~(\ref{eqn:normalization}) for our normalization convention.} \cite{Brodsky:2014yha}. It has been suggested
that in the massless limit one can choose the longitudinal confining strength to be independent of the quark mass to reproduce the
Gell-Mann-Oakes-Renner relation \cite{Gutsche:2012ez}. Our proposal shares some similarities with other proposals in the literature
\cite{Gutsche:2014oua, Trawinski:2014msa, Chabysheva:2012fe}.

\subsection{Running Coupling}

\begin{figure}[h]
\centering 
\includegraphics[width=0.5\textwidth]{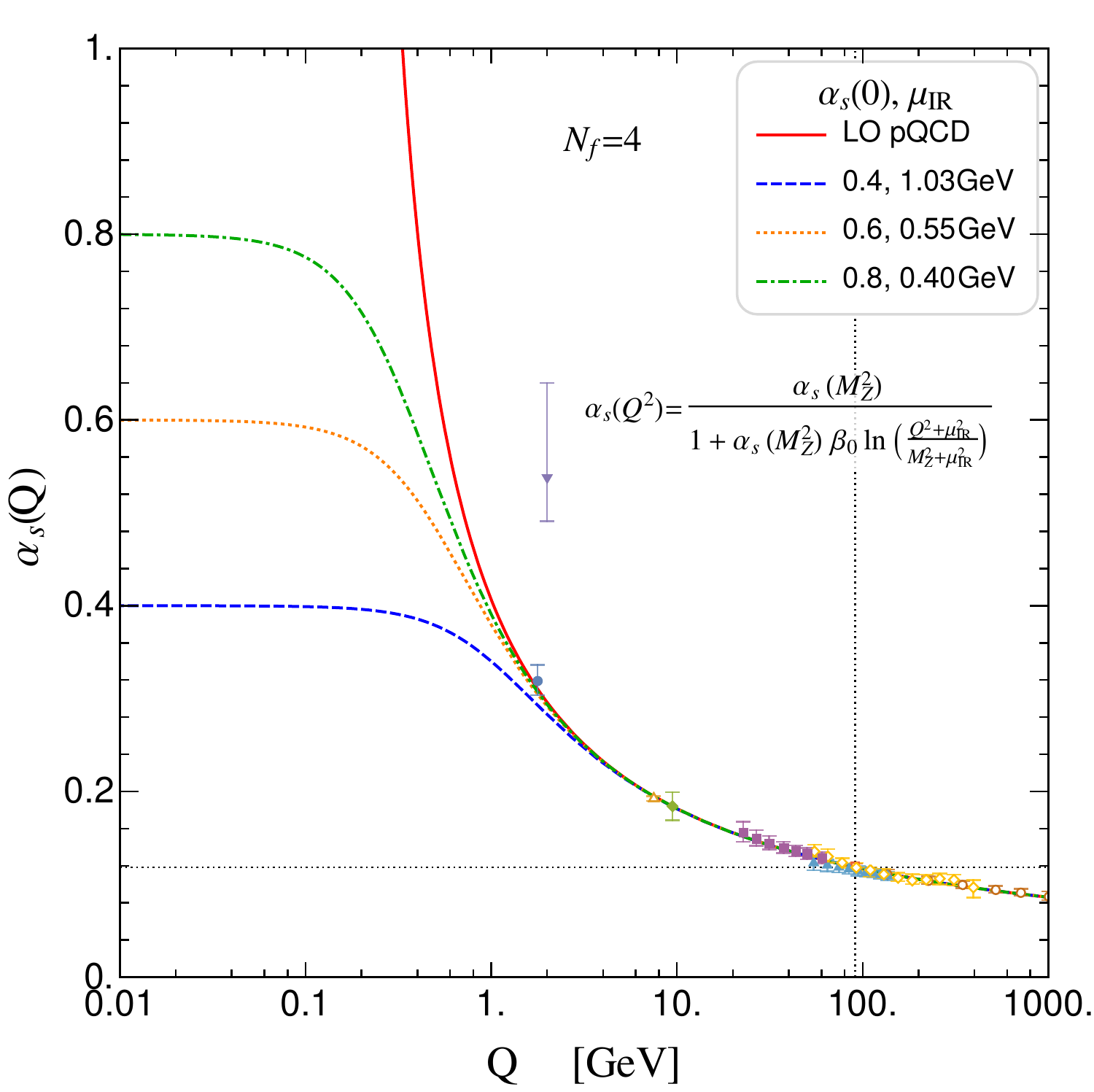}
\caption{The effective running coupling implemented in this work. Data points correspond to various experimental measurements. 
The vertical and horizontal lines mark the location of $M_\textsc{z}$ and $\alpha_s(M_\textsc{z}^2)$.}
\label{fig:runalf}
\end{figure}

As mentioned, we employ a running coupling based on the 1-loop pQCD. The running coupling is a function of the 4-momentum transfer squared
$Q^2=-q^2>0$ (see also Fig.~\ref{fig:runalf}), viz
\begin{equation}
\alpha_s(Q^2) = \frac{1}{\beta_0 \ln \big(Q^2/\Lambda^2 +\tau\big)} \triangleq 
\frac{\alpha_s(M_\textsc{z}^2)}{1+\alpha_s(M_\textsc{z}^2)\beta_0
\ln(\mu^2_\textsc{ir}+Q^2)/(\mu^2_\textsc{ir}+M^2_\textsc{z})},
\end{equation} 
where $\beta_0 = (33-2N_f)/(12\pi)$, with $N_f$ the number of quark flavors, $N_f = 4$ for charmonium and $N_f=5$ for bottomonium. 
A constant $\tau$ is introduced to avoid the pQCD IR catastrophe. Similar ans\"atze are widely adopted in the literature (e.g.
\cite{Maris:1999nt}). $\Lambda$ and constant $\tau$ are obtained by fixing the strong coupling at the Z-boson mass
$\alpha_s(M_\textsc{z}^2)=0.1183$ and at $Q=0$. 
In practice, we choose $\alpha_s(0) = 0.6$, corresponding to $\mu_\textsc{ir} = 0.55 \,\mathrm{GeV}$ for $N_f=4$. We find, however, the
spectra are not sensitive to the choice of $\alpha_s(0)$ within the range of $0.4\le
\alpha_s(0) \le 0.8$.

Introducing the evolution of the strong coupling implements asymptotic freedom for the one-gluon exchange through a natural dependence on
the covariant 4-momentum transfer $Q^2$. 
The use of the running coupling also serves to improve the UV asymptotics of the one-gluon exchange kernel. In our previous work
\cite{Li:2015zda}, we used a fixed coupling. The effective one-gluon exchange kernel, as derived from the leading-order effective
Hamiltonian approach, produces a divergent results, as is well known in the literature (e.g., Refs.~\cite{Krautgartner:1991xz,
Gubankova:1997mq,
ManginBrinet:2003nm}). This divergence is the result of the high momentum contribution from the spin non-flip part of the Hamiltonian matrix
elements. It can be easily seen from the power counting in transverse momenta. In Ref.~\cite{Li:2015zda}, we adopted a UV counterterm
proposed by Krautg\"artner, Pauli and W\"olz (KPW) \cite{Krautgartner:1991xz} (cf. Refs.~\cite{Glazek:1992bs, Trittmann:1997xz,
Wiecki:2014ola,
Lamm:2016djr}). However, the KPW counterterm is non-covariant, and introduces a major source of violation of the rotational symmetry that is
manifested in the spectrum. With asymptotic freedom, the UV divergence associated with the one-gluon exchange kernel is absent. Therefore,
the non-covariant KPW counterterm is not needed and we omit it in the present work. As we will see below, the rotational symmetry is
improved compared to the results of Ref.~\cite{Li:2015zda}. 

\section{Hamiltonian formalism}\label{sect:formalism}

\subsection{Eigenvalue Equation}
The mass spectrum and the wave functions are obtained from diagonalizing the effective light-front Hamiltonian operator (\ref{eqn:Heff}): 
\begin{equation}\label{eqn:eigenvalue_equation}
H_\text{eff} |\psi_h(P, j, m_j)\rangle = M^2_h |\psi_h(P, j, m_j)\rangle.
\end{equation}
where $P=(P^-, P^+, \vec P_\perp)$ is the 4-momentum of the particle; $j$ and $m_j$ are the particle's total angular momentum and the
magnetic projection, respectively.

The Fock space representation of quarkonium reads:
\begin{multline}\label{eqn:Fock_expansion}
 |\psi_h(P, j, m_j)\rangle = 
\sum_{s, \bar s}\int_0^1\frac{\dd x}{2x(1-x)} \int \frac{\dd^2 k_\perp}{(2\pi)^3}
\, \psi^{(m_j)}_{s\bar s/h}(\vec k_\perp, x) \\
\times \frac{1}{\sqrt{N_c}}\sum_{i=1}^{N_c} b^\dagger_{s{}i}\big(xP^+, \vec k_\perp+x\vec P_\perp\big) d^\dagger_{\bar s{}i}\big((1-x)P^+,
-\vec k_\perp+(1-x)\vec P_\perp\big) |0\rangle. 
\end{multline}
The coefficients of the expansion, $\psi^{(m_j)}_{s\bar s/h}(\vec k_\perp, x)$ are the valence sector LFWFs with $s$ ($\bar s$) representing
the spin of the quark (antiquark). The quark and anti-quark creation operators $b^\dagger$ and
$d^\dagger$ satisfy the canonical anti-commutation relations,
\begin{equation}\label{eqn:CCR}
 \big\{ b_{si}(p^+, \vec p_\perp), b_{s'i'}^\dagger(p'^+, \vec p'_\perp) \big\} 
= \big\{ d_{si}(p^+, \vec p_\perp), d_{s'i'}^\dagger(p'^+, \vec p'_\perp) \big\} 
= 2p^+(2\pi)^3\delta^3(p-p')\delta_{ss'}\delta_{ii'},
\end{equation}
where $\delta^3(p-p') \equiv \delta(p^+-p'^+)\delta^2(\vec p_\perp-\vec p'_\perp)$. We have kept only the $q\bar q$ sector while, in
principle, the $q\bar qg$ sector can be included by, e.g., a perturbative treatment \cite{Leitner:2010nx}.   
The hadron state vector can be orthonormalized according to the one-particle state [cf. Eq.~(\ref{eqn:CCR})]:
\begin{equation}\label{eqn:normalizaiton_of_state_vector}
 \langle \psi_h(P, j, m_j) | \psi_{h'}(P', j', m_j')\rangle = 2P^+ (2\pi)^3 \delta^3(P - P')
\delta_{jj'}\delta_{m_j,m_j'} \delta_{hh'},
\end{equation}
Then, the orthonormalization of the LFWFs reads, 
\begin{equation}\label{eqn:normalization}
 \sum_{s, \bar s} \int_0^1\frac{\dd x}{2x(1-x)} \int \frac{\dd^2  k_\perp}{(2\pi)^3} 
\psi^{(m_j')*}_{s \bar s/h'}(\vec k_\perp, x)\psi^{(m_j)}_{s \bar s/h}(\vec k_\perp, x)  = \delta_{hh'}\delta_{m_j,m_j'}.
\end{equation}
Note that different hadron states with the same quantum numbers, such as $J/\psi$ and $\psi'$, are also orthogonal. It is also useful to
introduce LFWFs in the transverse
coordinate space:
\begin{equation}
  \widetilde\psi_{s\bar s} (\vec r_\perp, x) \equiv \frac{1}{\sqrt{x(1-x)}}\int \frac{\dd^2 k_\perp}{(2\pi)^2}  e^{\imag \vec k_\perp \cdot
\vec r_\perp} \psi_{s \bar s} (\vec k_\perp, x).
\end{equation} 
with orthonormalization,
\begin{equation}
\sum_{s, \bar s} \int_0^1\frac{\dd x}{4\pi} \int \dd^2 r_\perp \, \widetilde\psi^{(m_j')*}_{s \bar s/h'}(\vec r_\perp, x)
\widetilde\psi^{(m_j)}_{s \bar s/h}(\vec r_\perp, x)  = \delta_{hh'}\delta_{m_j,m_j'}.
\end{equation}

Parity $\mathcal P$ is a dynamical symmetry on the light front, as it swaps light-front coordinate $x^-$ and light-front time $x^+$. The
mirror parity $m_P \equiv \mathcal R_x(\pi)\mathcal P$, which only flips one of the transverse spatial coordinates ($x^1$), survives as a
kinematical symmetry in light-front dynamics. The eigenvalue equations related to the mirror parity $\hat m_P$ and the charge conjugation
$\hat C$ are \cite{Krautgartner:1991xz, Trittmann:1997xz, Brodsky:2006ez, Li:2015zda}:
\begin{equation}\label{eqn:parity}
 \hat m_P |\psi_h(P, j, m_j)\rangle = (-i)^{2j}\mathsf{P}|\psi_h(\widetilde P, j, -m_j)\rangle, \quad \hat C |\psi_h(P, j, m_j)\rangle =
\mathsf C |\psi_{\bar h}(P, j,
m_j)\rangle. 
\end{equation}
Here $\mathsf P$ and $\mathsf C$ are the parity and charge conjugation quantum numbers, respectively; and $P=(P^-, P^+, P^1, P^2)$ is the
total 4-momentum of the particle, $\widetilde P = (P^-, P^+, -P^1, P^2)$. $\bar h$ represents the antiparticle of hadron $h$.

Particles are further classified by the eigenvalues of the intrinsic angular momenta $\{\vec{\mathcal J}^2,
\mathcal J_z \}$, viz
\begin{equation}
 \vec{\mathcal J}^2 |\psi_h(P, j, m_j)\rangle = j(j+1) |\psi_h(P, j, m_j)\rangle, \quad
 \mathcal J_z |\psi_h(P, j, m_j)\rangle = m_j |\psi_h(P, j, m_j)\rangle.
\end{equation}
On the light front, $\vec{\mathcal J}^2$ is dynamical and, in principle, it should be diagonalized simultaneously with the light-front
Hamiltonian operator $P^-$ to obtain the total angular momentum $j$ \cite{Brodsky:1997de}. Accordingly, in a truncated and regularized model
space, $\vec{\mathcal
J}^2$ may not commute with $P^-$, and the rotational symmetry is only approximate (see Fig.~\ref{fig:raw_spectrum}). To extract $j$,
we compute the mass eigenvalues from all $m_j$ sectors. We count the multiplicity of the nearly-degenerate mass eigenstates with the further
help of the mirror parity, charge conjugation and other relevant quantities\footnote{For example, the decay constants.}. For this
scheme to succeed, the degeneracies have to be observed in the results with sufficient accuracy to resolve ambiguities.

\subsection{Basis Representation}\label{sec:3.1}

\begin{figure}
 \centering 
 \subfloat[\ \label{fig:basis_functions_a}]{\includegraphics[width=0.48\textwidth]{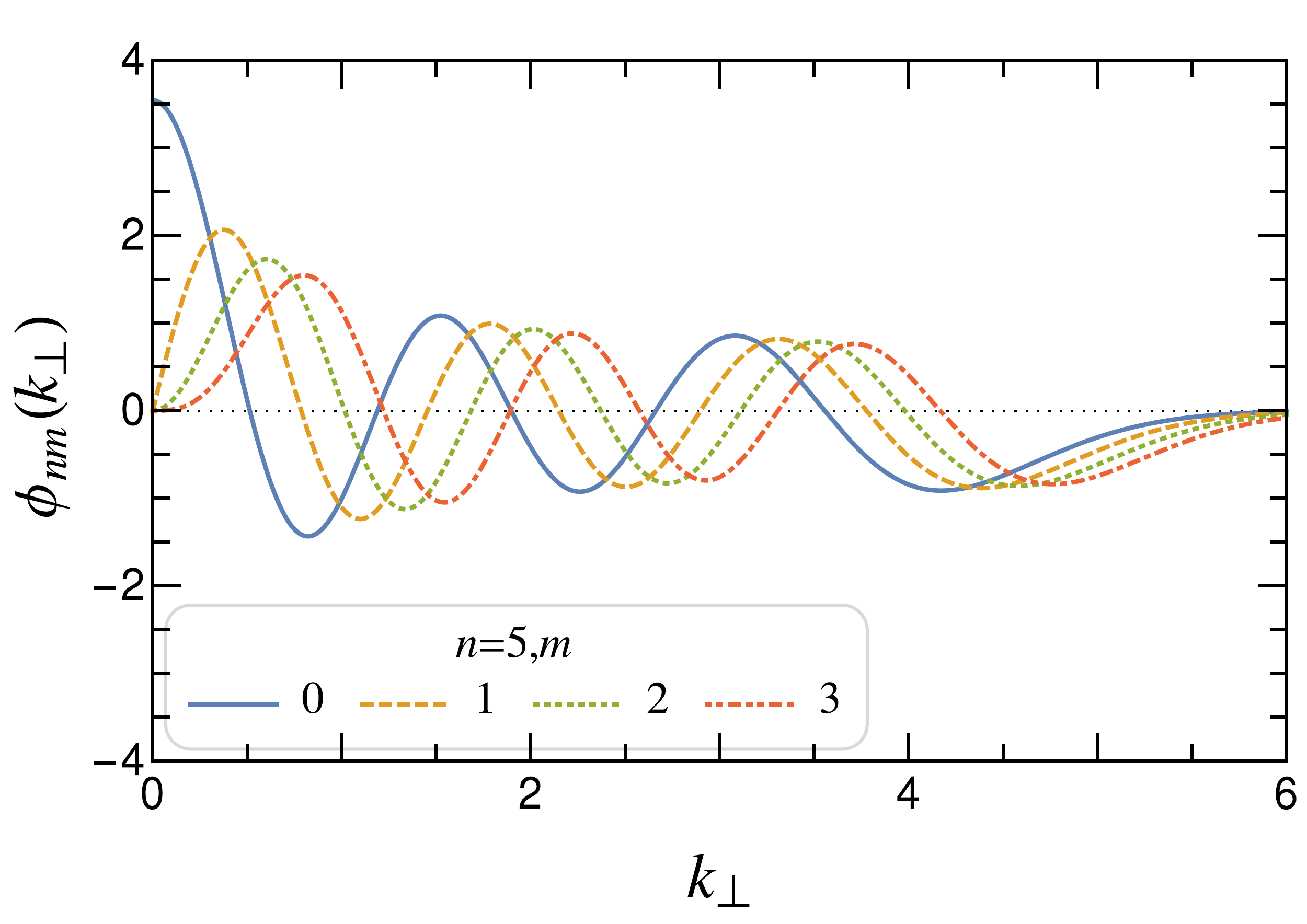}} \quad 
 \subfloat[\ \label{fig:basis_functions_b}]{\includegraphics[width=0.48\textwidth]{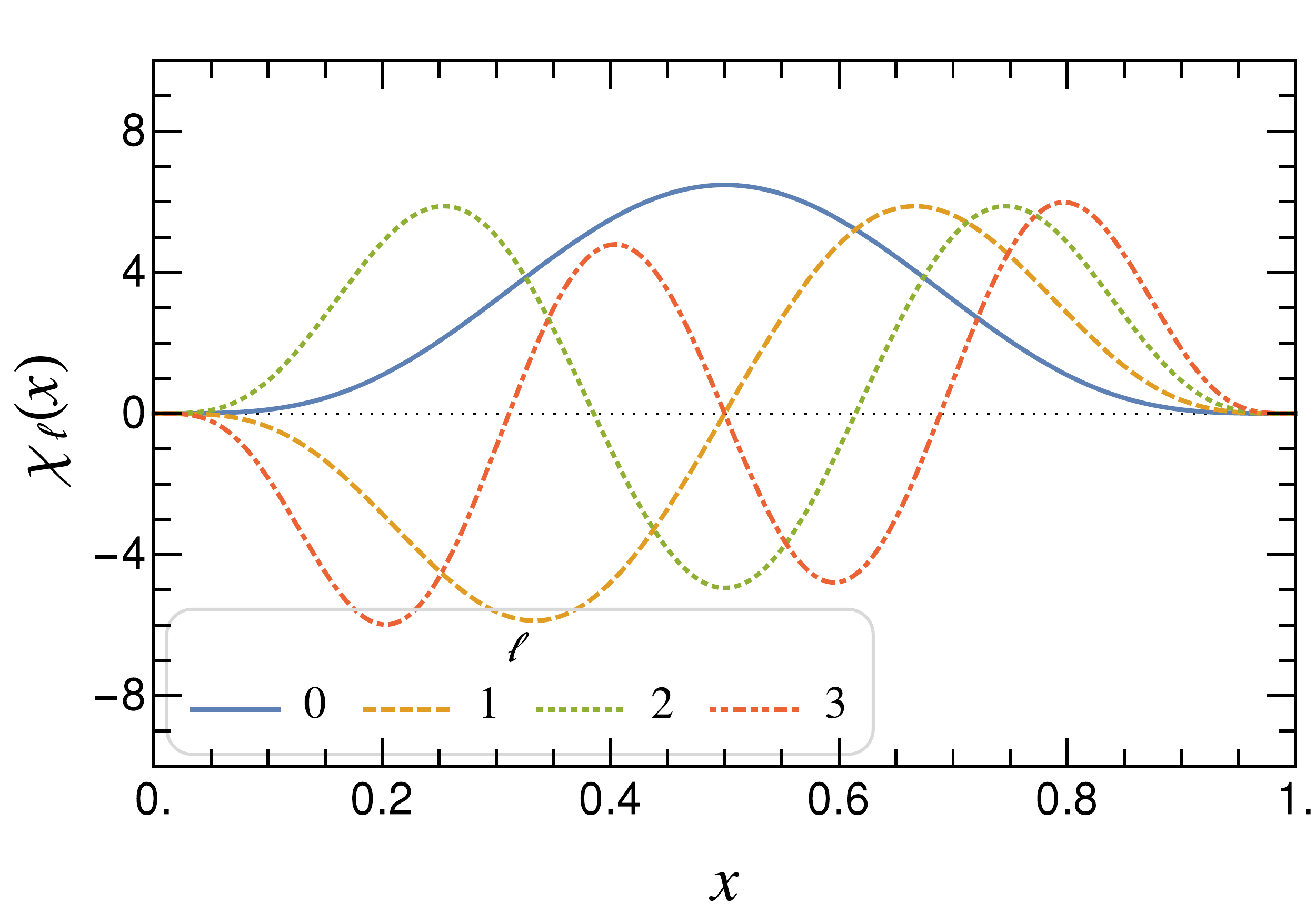}}
 \caption{ 
 \textit{Left panel}: the transverse basis function $\phi_{nm}(\vec k_\perp; b)$ at $b=1$, $n=5$, $\arg \vec k_\perp=0$;
 \textit{Right panel}: the longitudinal basis function $\chi_\ell(x; \alpha,\beta)$ at $\alpha=\beta=16$.}
 \label{fig:basis_functions}
\end{figure}

The eigenvalue equation (\ref{eqn:eigenvalue_equation}) can be solved in a basis function approach \cite{Vary:2009gt, Li:2015zda}. The basis
function approach is particularly advantageous for the present model with the holographic confining potential, since, in the absence of the
one-gluon exchange term, it can be diagonalized analytically. On the other hand, the confining interactions in momentum space are
highly singular.
The solutions can be expressed in terms of the analytic functions $\phi_{nm}$ and $\chi_l$. For the transverse direction, we have (see
Fig.~\ref{fig:basis_functions_a}):
\begin{equation}
 \phi_{nm}(\vec q_\perp; b) = b^{-1} \sqrt{\frac{4\pi n!}{(n+|m|)!}} \bigg(\frac{q_\perp}{b}\bigg)^{|m|}
\exp\big(-q^2_\perp/(2b^2)\big) L_n^{|m|}(q^2_\perp/b^2) \exp\big(\imag m \theta_q),  
\end{equation}
where $\vec q_\perp \triangleq \vec k_\perp/\sqrt{x(1-x)}$, $q_\perp = |\vec q_\perp|$, $\theta_q = \arg \vec q_\perp$. $L_n^a(z)$ is the
associated Laguerre polynomial. $b$ is the harmonic oscillator (HO) basis parameter in mass dimension. Following Ref.~\cite{Li:2015zda}, we
choose $b\equiv \kappa$ to match the confining strength. For simplicity, we will often omit the label $b$ though it is implicit throughout.
In the longitudinal direction, we have (see Fig.~\ref{fig:basis_functions_b}):
\begin{equation}
 \chi_l(x; \alpha,\beta) =
\sqrt{4\pi(2l+\alpha+\beta+1)}\sqrt{\frac{\Gamma(l+1)\Gamma(l+\alpha+\beta+1)}{\Gamma(l+\alpha+1)\Gamma(l+\beta+1)}}
x^{\half[\beta]}(1-x)^{\half[\alpha]} P^{(\alpha,\beta)}_l(2x-1).
\end{equation}
Here $P_l^{(\alpha, \beta)}(z)$ is the Jacobi polynomial. $\alpha$ and $\beta$ are dimensionless basis parameters. In the model, they are 
$\alpha = 2m_{\bar q}(m_q+m_{\bar q})/\kappa^2$, $\beta = 2m_{q}(m_q+m_{\bar q})/\kappa^2$. Again, we will drop the explicit dependence 
on $\alpha$ or $\beta$ from now on.

In the presence of the one-gluon exchange term, we use these analytic functions as a basis to expand the LFWFs in, 
\begin{equation}\label{eqn:basis_representation}
 \psi_{ss'/h}(\vec k_\perp, x) = \sum_{n, m, l} \psi_h(n, m, l, s, s') \, \phi_{nm}(\vec k_\perp/\sqrt{x(1-x)}) \chi_l(x).
\end{equation}
Here the coefficients $ \psi_h(n, m, l, s, s')$ are obtained from diagonalization. The basis is constructed to conserve the magnetic
projection of the total angular momentum: $ m_j = m + s + s'$.

Performing a 2D Fourier transformation gives the LFWFs in coordinate space. The Fourier transformation of a HO function is a HO function
with a relative phase, which simplifies the expression greatly.
\begin{equation}
 \widetilde \psi_{ss'/h}(\vec r_\perp, x) = \sqrt{x(1-x)} \sum_{n, m, l} \psi_h(n, m, l, s, s') \, \widetilde \phi_{nm}(\sqrt{x(1-x)} \vec
r_\perp) \chi_l(x).
\end{equation}
Here $\widetilde \phi_{nm}$ is the 2D HO in coordinate space:
\begin{equation}
 \widetilde \phi_{nm}(\vec \rho_\perp; b^{-1}) = b \sqrt{\frac{n!}{\pi(n+|m|)!}} (b\rho_\perp)^{|m|} \exp\big(-b^2\rho^2_\perp/2\big)
L_n^{|m|}(b^2\rho_\perp^2) \exp\big[ \imag m\theta_\rho + \imag \pi (n+|m|/2) \big].
\end{equation}

In practical calculations, the basis is truncated and wave functions are obtained in the basis expansion. Following
Refs.~\cite{Vary:2009gt, Wiecki:2014ola, Li:2015zda}, we truncate the transverse and the longitudinal bases separately by their energies:
\begin{equation}
 2n + |m| + 1 \le N_\mathrm{max}, \quad 0 \le l \le L_{\mathrm{max}}.
\end{equation}
As such, the $N_{\max}$-truncation provides a natural pair of UV and IR cutoffs: $\Lambda_\textsc{uv} \simeq b\sqrt{N_{\max}}$,
$\lambda_\textsc{ir} \simeq b/\sqrt{N_{\max}}$, where $b=\kappa$ is the oscillator basis energy scale parameter. $L_{\max}$ represents the
resolution of the basis in the longitudinal direction. Namely, the basis cannot resolve physics at: $\Delta x\lesssim L_{\max}^{-1}$
\cite{Wiecki:2014ola}. The complete basis is reached by taking $N_{\max}\to\infty, L_{\max} \to \infty$.

The eigenvalues of the parity and charge conjugation operators can be  extracted from the basis representation of the LFWFs as
\cite{Li:2015zda},
\begin{align}
 (-i)^{2j}\mathsf P =\,& \langle\psi_{-m_j}|\hat m_P | \psi_{m_j}\rangle = \sum_{n,m,l,s,\bar s} (-1)^m \psi_{-m_j}^*(n, -m, l, -s, -\bar s)
\psi_{m_j}(n, m, l, s, \bar s). \label{eqn:LFP}\\
 \mathsf C =\,& \langle\psi_{m_j}|\hat C | \psi_{m_j}\rangle = \sum_{n,m,l,s,\bar s} (-1)^{m+l+1} \psi_{m_j}^*(n, m, l, \bar s, s)
\psi_{m_j}(n, m, l, s, \bar s). \label{eqn:C}
\end{align}

\subsection{Generalizing Light-Front Holography}

Before proceeding to the full diagonalization, it is worth looking at the results without the one-gluon exchange, where the solutions are
analytical. The mass eigenvalues are:
\begin{equation}
 M^2_{n,m,l} = (m_q+m_{\bar q})^2 + 2\kappa^2(2n+|m|+l+1) + \frac{\kappa^4}{(m_q+m_{\bar q})^2}l(l+1).
\end{equation}
\emph{Here $l$ is the longitudinal quantum number, not the orbital angular momentum}. The corresponding wave functions are:
\begin{equation}
 \psi_{nml}(\vec k_\perp, x) = \phi_{nm}(\vec k_\perp/\sqrt{x(1-x)}) \chi_l(x).
\end{equation}

\begin{figure}
 \centering 
  \includegraphics[width=0.5\textwidth]{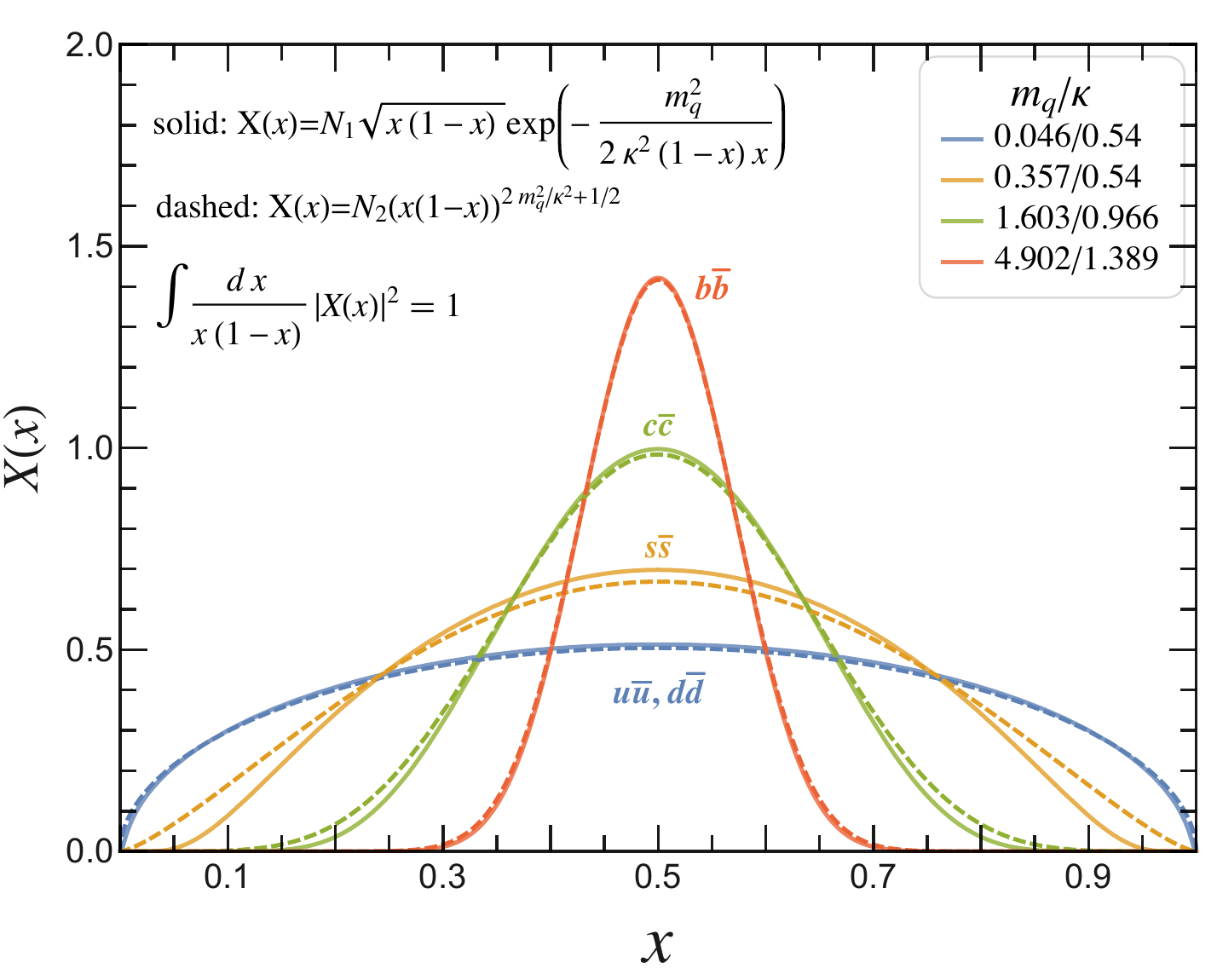}
 \caption{ 
 Comparison of the ground-state longitudinal wave functions obtained from the invariant mass ansatz:
$N_1\exp\big(-\frac{m^2_q}{2\kappa^2x(1-x)}\big)$ and from BLFQ: $N_2 \big( x(1-x)\big)^{2m^2_q/\kappa^2}$. 
We convert the wave functions to those of the Brodsky-de T\'eramond convention \cite{Brodsky:2014yha} by including a factor
$\sqrt{x(1-x)}$ [cf. Eq.~({\ref{eqn:normalization}})]. 
Quark mass $m_q$ and confining strength $\kappa$ are taken from the fits of Ref.~\cite{Brodsky:2014yha} and this work
(Sect.~\ref{sect:numerical_results}).
}
 \label{fig:longitudinal_basis_functions}
\end{figure}

States may be identified according to their mass spectrum with the help of parity $\mathsf P$ and charge conjugation $\mathsf C$. 
The quarkonium ground state (1S) is identified with $n=0, m=0, l=0$:
\begin{equation}
 \psi_\mathrm{gs}(\vec k_\perp, x) = N \exp\big[ -\vec k^2_\perp/(2\kappa^2 x(1-x)) \big] \big( x(1-x)\big)^{2m^2_q/\kappa^2}.
\end{equation}
In the literature, a commonly-used way to incorporate quark masses in the AdS/QCD wave function is through the invariant mass ansatz (IMA)
\cite{Brodsky:2008pg}, viz,
\begin{equation}
N \exp\big[ -\vec k^2_\perp/(2\kappa^2 x(1-x)) \big] \; \to \; N' \exp\big[ -(\vec k^2_\perp+m_q^2)/(2\kappa^2 x(1-x)) \big]. 
\end{equation}
Figure~\ref{fig:longitudinal_basis_functions} compares the purely longitudinal part of our ground-state wave function with that of the IMA
wave function. Our longitudinal wave function becomes almost identical to the IMA wave function in both the chiral limit and the heavy
quark limit, except near the
endpoints. This reflects the fact that rotational symmetry is restored in the non-relativistic limit with our choice of longitudinal basis
functions. 

The first excited state (1P) is identified with $n=0, m=\pm1, l=0$ or $n=0, m=0, l=1$, noting that for heavy quarkonium, the term
$\kappa^4/(m_q+m_{\bar q})^2 l(l+1)$ is small comparing to the remaining terms. There are four 1P states: $\chi_0$ ($0^{++}$), $\chi_1$
($1^{++}$), $\chi_2$ ($2^{++}$) and $h$ ($1^{+-}$). Let us focus on $h$ and restrict the discussion to $m_j=0$. From Eq.~(\ref{eqn:C}), we
conclude: $-1 = \mathsf C = (-1)^{m+l+1}(-1)^{s+1}$, where $s$ is the total spin, viz $s=0$ for singlet and $s=1$ for triplet. Apparently,
for both sets of quantum numbers ($m=\pm1, l=0$ or $m=0, l=1$), $s=0$. From Eq.~(\ref{eqn:LFP}), $-1 = (-1)^j\mathsf P =
(-1)^{m}(-1)^{s+1}$, implying $m=0$. Therefore, the correct quantum numbers for $h$ meson ($1^{+-}$) are $n=0, m=0, l=1$ with a singlet spin
configuration, which is consistent with the non-relativistic quantum number assignment $1{\,}^{1}\!P_0$. Note that the orbital motion
is excited through the longitudinal direction but not the transverse direction. This cannot be obtained from IMA\footnote{In the literature,
the longitudinal excitations are typically obtained from modeling the spin structure via the spinor wave function $\bar u \Gamma v$.
However, the longitudinal profile of the spinor wave function is qualitatively different from the holographic wave function.
}.

\section{Numerical Results}\label{sect:numerical_results}

\begin{table}[h]
\caption{Summary of the model parameters (see text). 
}\label{tab:model_parametersI}
 \centering
\begin{tabular}{ccc ccc ccc c} 

\toprule 

  & $N_f$ & $\alpha_s(0)$ & $\mu_\text{g}$ (GeV) & $\kappa$ (GeV) & $m_q$ (GeV) & rms (MeV) & $\overline{\delta_jM}$ (MeV) & $N_\text{exp}$
&
$N_{\max}=L_{\max}$  \\ 

\colrule

\multirow{1}{*}{$c\bar c$}  & \multirow{1}{*}{4} & 0.6 & 0.02 & 0.966 & 1.603 & 31{\qquad\;\;}   &
17{\qquad\;\;}  & 8  & 32 \\ 

\multirow{1}{*}{$b\bar b$}  & \multirow{1}{*}{5} & 0.6 & 0.02 & 1.389 & 4.902 & 38{\qquad\;\;}  &
8{\qquad\;\;}   & 14 & 32 \\ 

\botrule  
\end{tabular}
\end{table}

We apply the model to heavy quarkonia (charmonium and bottomonium), where the quark masses are large and the radiative corrections are
negligible. Therefore these are ideal systems to test our model. The model parameters are summarized in Table~\ref{tab:model_parametersI}. 

As mentioned, we fixed $\alpha_s(0)=0.6$. For fixed $N_{\max}$ and $L_{\max}$, we use experimental data to fit the confining strength
$\kappa$ and the effective quark mass $m_q$ ($m_c$ and $m_b$) using the mass eigenvalues in the $m_j=0$ sector. We employ the experimental
values, compiled by the Particle Data Group (PDG) \cite{Agashe:2014kda}, below the open charm or open bottom threshold. 
We also introduced a small mass parameter $\mu_g = 0.02 \,\mathrm{GeV}$ to regularize the integrable Coulomb singularity in the energy
denominator and to avoid numerical instability\footnote{Our numerical method is designed such that no singularity is encountered in the
actual calculation. Nevertheless, we introduced this parameter, smaller than all other energy scales, to further tame the integrable
singularity. }. As has been shown in previous work with fixed coupling, the mass eigenvalues are converged with respect to $\mu_g \to 0$
within the numerical precision.

\begin{figure}
 \centering 
 \includegraphics[width=0.6\textwidth]{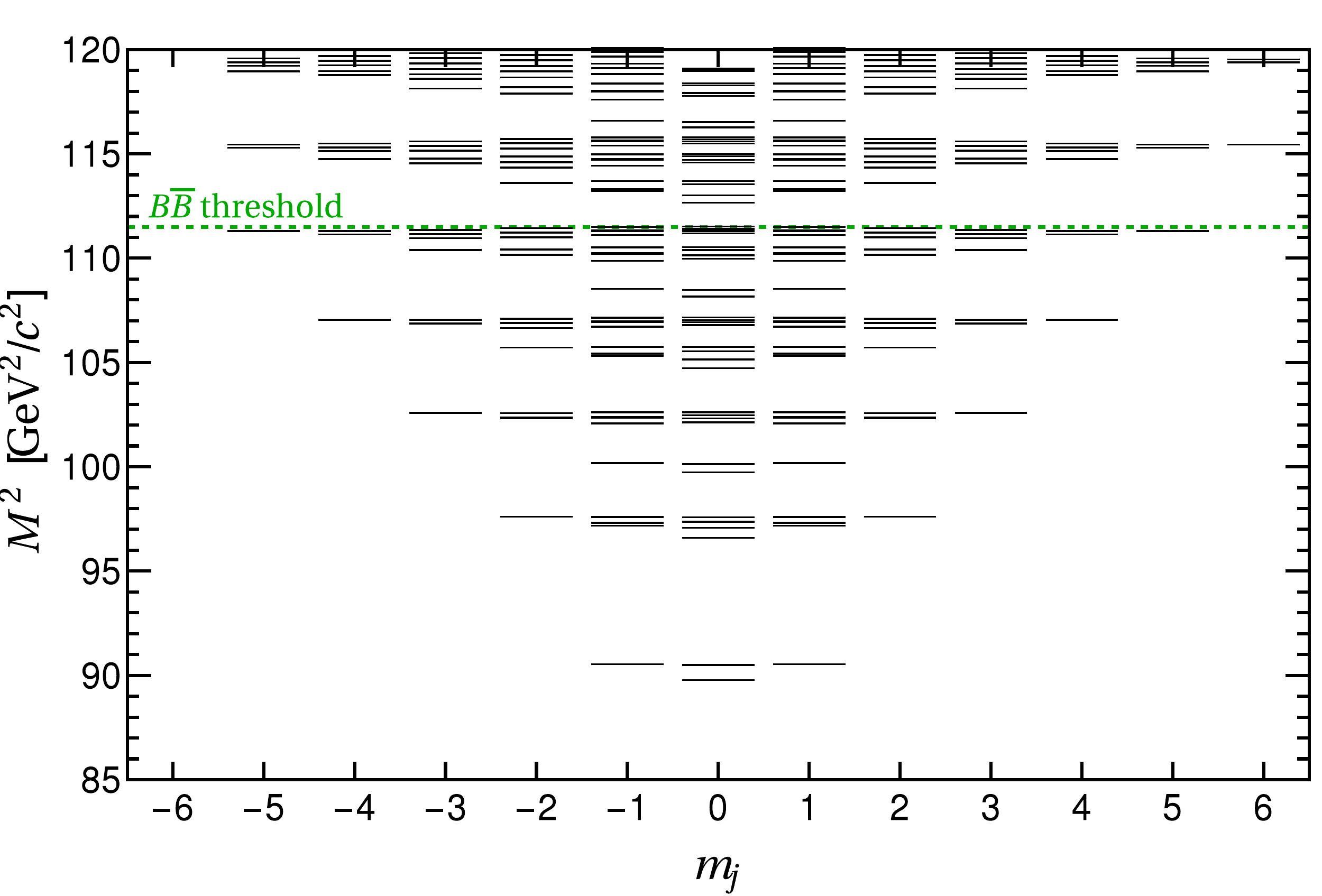}
 \caption{A representative bottomonium mass spectrum obtained by diagonalizing the light cone
Hamiltonian within various $m_j$ sectors at $N_{\max} = L_{\max} = 32$. Even though the rotational symmetry 
is not exact, the approximate degeneracies are sufficient to extract $j$. States with the same orbital
angular momentum $\ell$ tend to cluster, as expected from the non-relativistic quark model, even though 
$\ell$ is not a good quantum number, which is also helpful for identifying states.
}
 \label{fig:raw_spectrum}
\end{figure}

The effective Hamiltonian (\ref{eqn:Heff}) is diagonalized for various $m_j$ sectors. Fig.~\ref{fig:raw_spectrum} shows a representative
spectrum as a function of $m_j$. The spectrum is symmetric with respect to $\pm m_j$, a consequence of the mirror parity symmetry
(\ref{eqn:parity}). The discrete quantum numbers $m_P = (-i)^{2j}\mathsf P$ and $\mathsf C$ are computed to help identify states as
mentioned. Total
spin $\langle\vec s^2\rangle=s(s+1)$ as an approximate quantum number is also exploited. 
States with the same $j$ but different $m_j$'s are not exactly degenerate owing to the violation of the rotational symmetry. As is seen in 
Fig.~\ref{fig:raw_spectrum}, the approximate degeneracies are easily visible, at least for low-lying states. So the multiplicities, together
with $m_P$, $\mathsf C$, $s$ and the constraints: 
\begin{equation}
|\ell - s| \le j \le \ell+s, \quad \mathsf P=(-1)^{\ell+1}, \quad \mathsf C = (-1)^{\ell+s},
\end{equation}
can be employed to deduce the full set of quantum numbers $n\,{}^{2s+1}\!\ell_j$ or $j^{\mathsf P\mathsf C}$, where $\ell$ is the total
orbital angular
momentum, $n$ the radial quantum number. We also cross-check the state identification with the decay constants and the wave functions
themselves (see Sect.~\ref{sect:lfwfs}).

\subsection{Spectroscopy}\label{sect:spectroscopy}

\begin{figure}
\centering 
\includegraphics[width=0.45\textwidth]{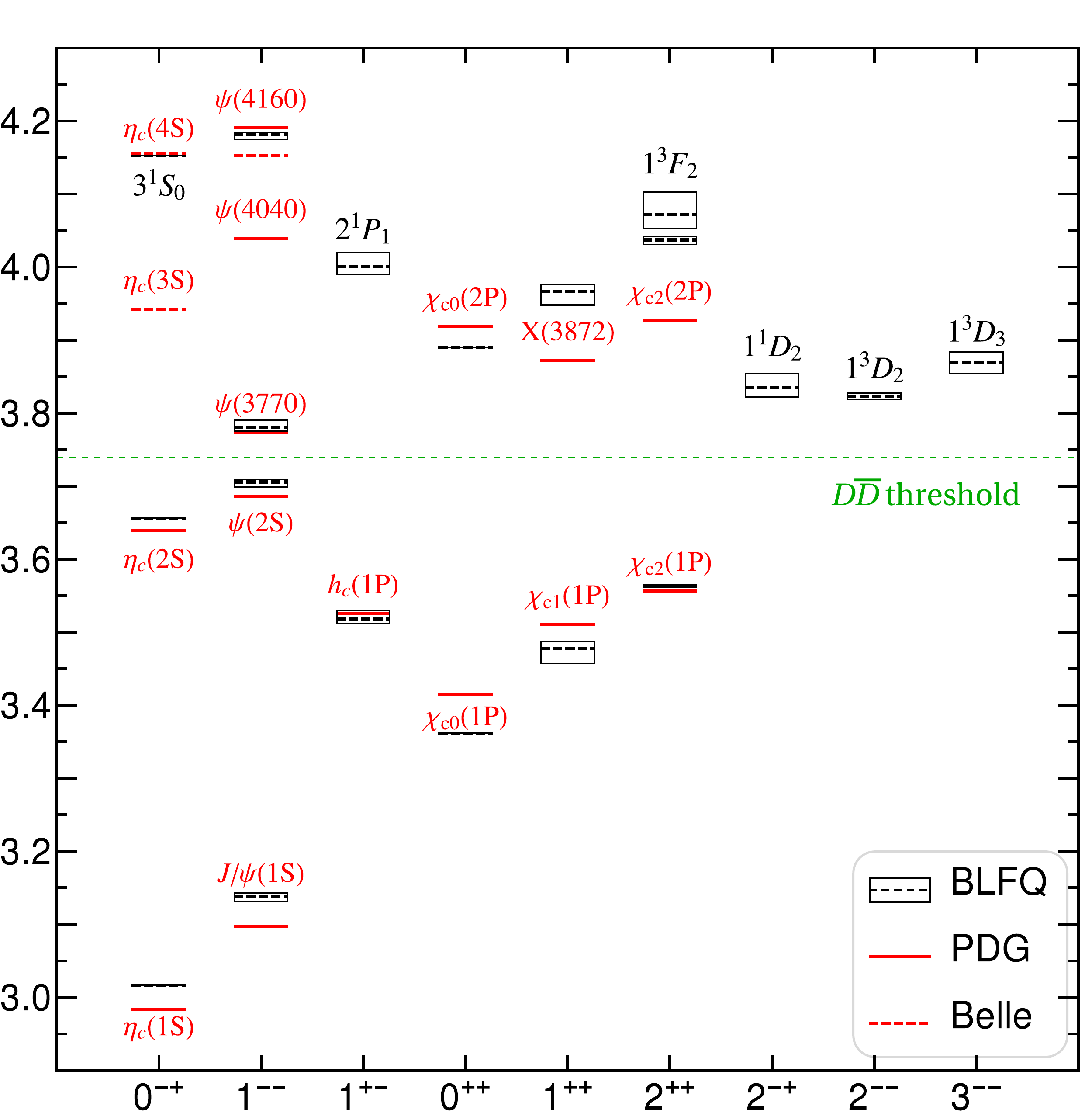}
\includegraphics[width=0.46\textwidth]{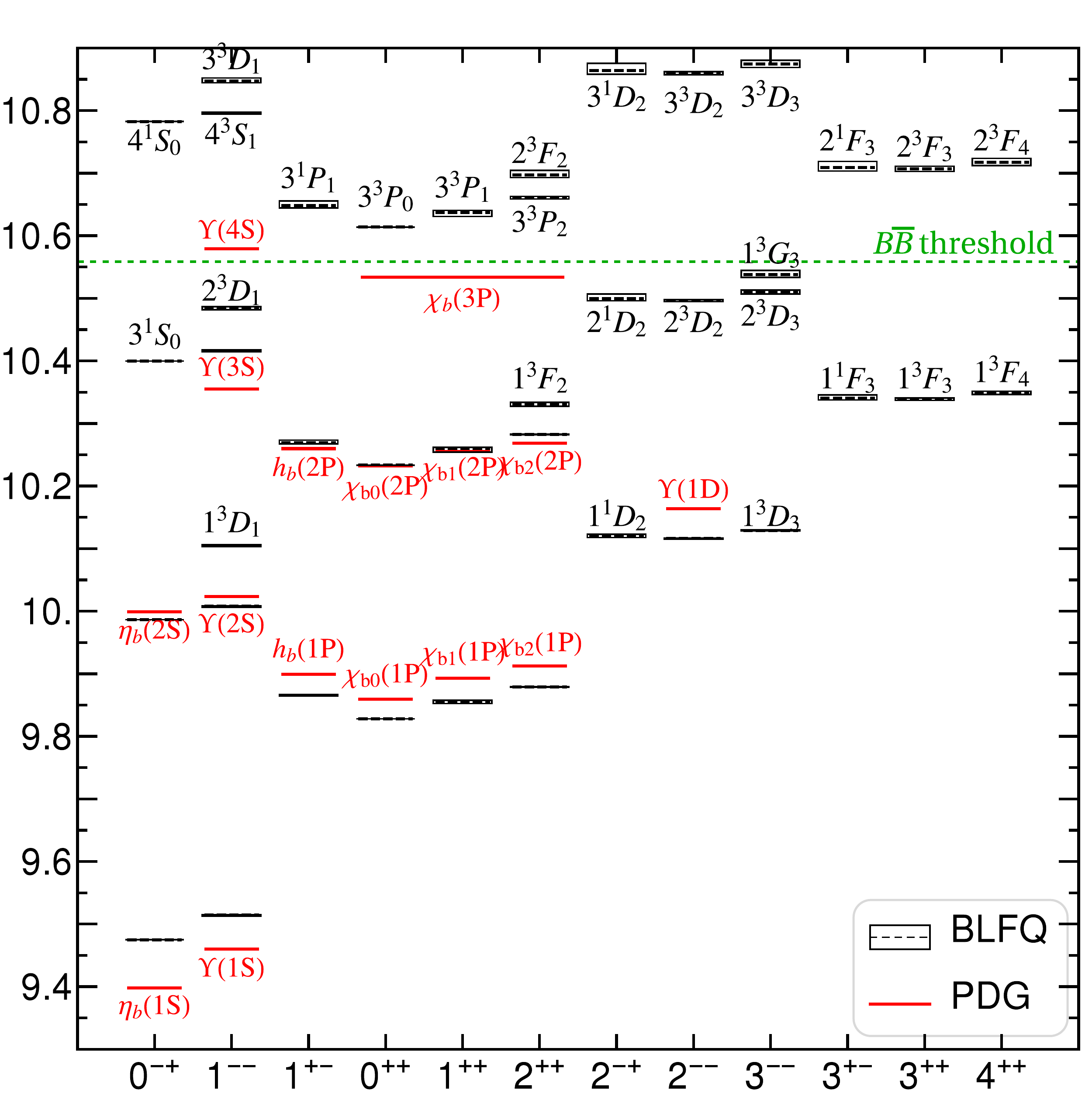}
\caption{The reconstructed charmonium (\textit{left panel}) and bottomonium (\textit{right panel}) spectra at $N_{\max}=L_{\max}=32$. 
The horizontal and vertical axises are $j^{\mathsf P\mathsf C}$ and invariant mass in GeV, respectively. 
Model parameters are listed
in Table~\ref{tab:model_parametersI}. Calculated states are marked by boxes to represent the spread of the mass eigenvalues in $m_j$ owing
to violation
of the rotational symmetry (see text). The mean mass spreads, i.e. the average heights of the boxes, are 17 MeV and 8 MeV for charmonium and
bottomonium, respectively. The r.m.s. deviations of the masses from the PDG values are 31 MeV and 38 MeV for charmonium and bottomonium,
respectively. See text for details.
}
\label{fig:spectroscopy}
\end{figure}

The reconstructed spectra at $N_{\max}=L_{\max}=32$ are presented in Fig.~\ref{fig:spectroscopy}. In these figures, we use boxes to indicate
the spreads of the mass eigenvalues from different $m_j$. The mean values, marked by dashed bars, are defined as:
\begin{equation}
 \overline M \equiv \sqrt{\frac{M^2_{-j} + M^2_{1-j} + \cdots + M^2_j}{2j+1}},
\end{equation}
where $M_{m_j}$ is the mass eigenvalue associated with the magnetic projection $m_j$. This definition is motivated by the covariant
light-front analysis of relativistic bound states in Refs.~\cite{Carbonell:1998rj, ManginBrinet:2003nm}. On the other hand, the mass spreads
 $\delta_j M \equiv \max M_{m_j} - \min M_{m_j}$ measure the violation of the rotational symmetry. We also introduce the mean spread:
\begin{equation}
 \overline{\delta_jM} \equiv \sqrt{ \frac{1}{N_h} \sum_{h}^{j\ne0} (\delta_jM_h)^2 }.\qquad \Big(N_h\equiv\sum_h^{j\ne0} 1\Big)
\end{equation}
For charmonium (bottomonium) states evaluated by PDG below the threshold, the mean mass spread is 17 MeV (8 MeV), improving 
our previous results \cite{Li:2015zda} by a factor of $\sim$3 ($\sim$2). More comparison between the results of this work and those of
Ref.~\cite{Li:2015zda} is collected in Table~\ref{tab:improvements}.

\begin{table}
\centering 
\caption{Comparison of differences between fits and PDG experimental data between results of Ref.~\cite{Li:2015zda} and those presented
here. $\delta M_{c\bar c}$ is the rms mass deviation for charmonium from the PDG data. $\overline{\delta_j M}_{c\bar c}$ is the mean mass
spread for charmonium. ``fix-$\alpha_s$ (refitted)'' improves the bottomonium fits by $\sim$10 MeV.} 
\label{tab:improvements}
\begin{tabular}{l|ccccc}
 \toprule 
   & $\overline{\delta_j M}_{c\bar c}$ & $\delta M_{c\bar c}$ (rms) & $\overline{\delta_j M}_{b\bar b}$ & $\delta M_{b\bar b}$ (rms) &
$N_{\max} = L_{\max}$ \\
 \colrule 
  fix-$\alpha_s$ \cite{Li:2015zda} & 49 MeV & 52 MeV & 17 MeV & *58 MeV & 24 \\
  fix-$\alpha_s$ (refitted) & --- & --- & 15 MeV & 48 MeV & 24 \\
  running-$\alpha_s$ & 17 MeV & 31 MeV & 7 MeV & 39 MeV & 24 \\
  running-$\alpha_s$ & 17 MeV & 31 MeV & 8 MeV & 38 MeV & 32 \\
 \botrule
\end{tabular}\\
{* In Ref.~\cite{Li:2015zda}, this is misquoted as 50 MeV.}
\end{table}

\begin{figure}
\centering 
\includegraphics[width=0.48\textwidth]{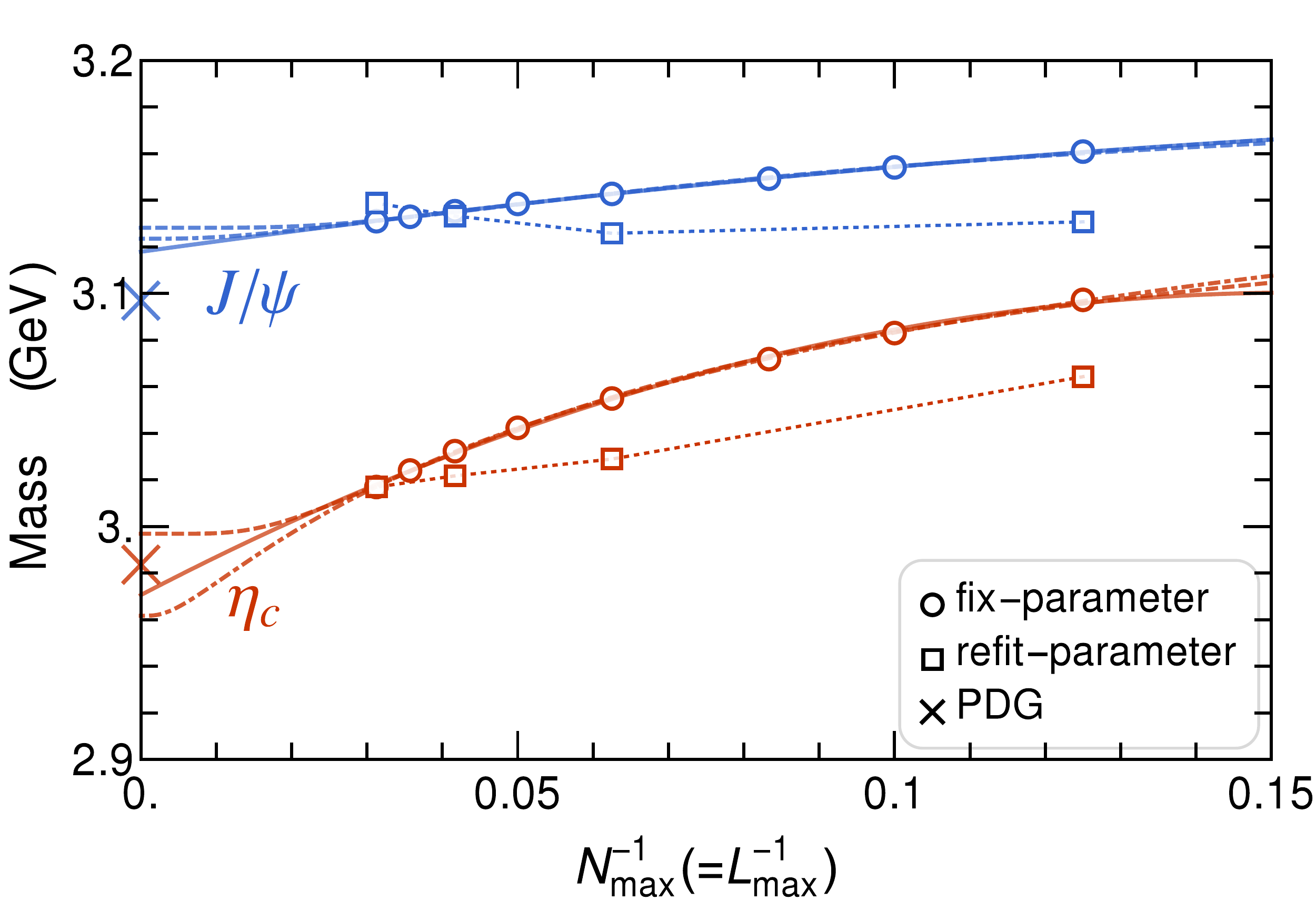}
\includegraphics[width=0.495\textwidth]{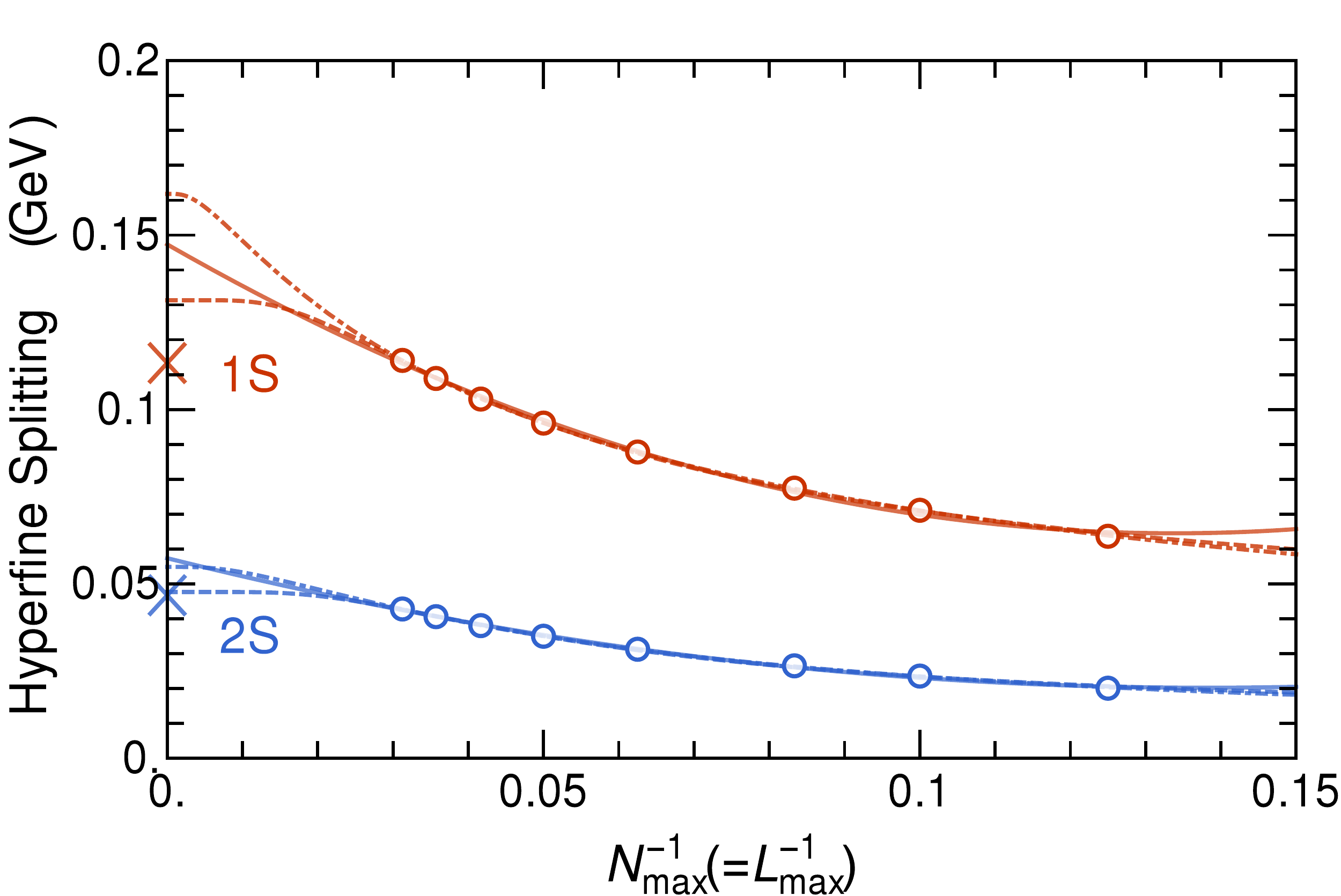}
\caption{The $N_{\max}$ convergence. %
The \textit{left panel} compares the $J/\psi$ and $\eta_c$ mass eigenvalues as a function of $N_{\max}^{-1}$ 
($N_{\max}=L_{\max}$, $m_j=0$) for fixed and refitted parameters. For the former (``fix-parameter''), parameters are the same for all
$N_{\max}$, and are chosen to be the fitted values at $N_{\max}=32$. For the latter (``refit-parameter''),  
parameters are refitted for each $N_{\max}$.
The \textit{right panel} shows the hyperfine splittings $M_{\psi(nS)}-M_{\eta_c(nS)}$ as a function of $N_{\max}^{-1}$
($N_{\max}=L_{\max}$, $m_j=0$) with 
fixed parameters.
The PDG values are marked as crosses. In both figures, different fitting functions, $a+b/N_{\max}+c/N_{\max}^2$ (solid), 
$a+b \exp(-c N_{\max})$ (dashed), $a+b \exp(-c \sqrt{N_{\max}})$ (dot-dashed), are shown for the fix-parameter results.
The refit-parameter results are simply connected by a straight line segments.
}
\label{fig:conv}
\end{figure}

Our light-front Hamiltonian approach yields states with high angular and radial excitations, which are not easily accessible in some other
methods. No exotic quantum numbers emerge from our calculation, as is expected from the two-body truncation. In bottomonium, predictions
are made for various states below the $B\overline B$ threshold, as also predicted in other approaches (e.g.,
\cite{Crater:2010fc, McNeile:2012qf, Hilger:2014nma}).
The quality of the spectra can be measured by the root mean squared (r.m.s.) deviation from the experimentally measured values.
For charmonium (bottomonium), the r.m.s. mass deviation is 31 MeV (38 MeV), improving the fixed coupling results \cite{Li:2015zda} by as
much as $\sim$40\% ($\sim$20\%). See Table~\ref{tab:improvements} for further comparisons. Our spectroscopy is competitive with those
obtained from other relativistic models \cite{Crater:2010fc, McNeile:2012qf, Hilger:2014nma, Leitao:2016bqq}. 
Not only are the mass spectra improved, the spread of the mass eigenvalues $\overline{\delta_j M}$ due to the violation of rotational
symmetry, is also significantly reduced as mentioned. 
A related issue is the quenching of the hyperfine splitting found within the fixed coupling results. With the running coupling, this issue
is resolved and the hyperfine splittings are consistent with the experimental values, as shown in Fig.~\ref{fig:spectroscopy}
(cf.~Fig.~\ref{fig:conv}). Therefore, the violation of the rotational symmetry is significantly reduced.

Figure~\ref{fig:conv} shows the trends of the charmonium mass eigenvalues as functions of $N_{\max}^{-1}$ (with $N_{\max}=L_{\max}$). The
left panel presents the convergence trends of the ground-state masses ($\eta_c$ and
$J/\psi$). The right panel presents the convergence trends for the hyperfine splittings between 1S ($M_{J/\psi}-M_{\eta_c}$) and 2S
($M_{\psi'}-M_{\eta'_c}$) states. Two sets of parameters are used: the fix-parameter results use model parameters from the
$N_{\max}=L_{\max}=32$ fit; the refit-parameter calculation refits the model parameters for each $N_{\max}=L_{\max}$. Smooth extrapolations
are made using three types of functions: $a+b/N_{\max}+c/N_{\max}^2$ (solid), $a+b \exp(-c N_{\max})$ (dashed), $a+b \exp(-c
\sqrt{N_{\max}})$ (dot-dashed). Both hyperfine splittings, 1S and 2S, show reasonable convergence in the complete basis limit
($N_{\max}^{-1}=L_{\max}^{-1}=0$). 

Table~\ref{tab:model_parametersII} compares the spectroscopy obtained from different $N_{\max} = L_{\max}$ fits. While results from 
different $N_{\max} = L_{\max}$ are well converged, the r.m.s. deviation decreases as $N_{\max}=L_{\max}$ increases. In the
present work, we adopt $N_{\max}=L_{\max}=32$ for our presented results, unless otherwise specified.

 \begin{table}
 \caption{
 Model sensitivity with respect to the basis size $N_{\max}=L_{\max}$. 
 The model parameters fits and the r.m.s. deviations are well converged 
as $N_{\max}=L_{\max}$ increases.
 }\label{tab:model_parametersII}
  \centering
 \begin{tabular}{c ccc ccc ccc} 
 
 \toprule 
 
  & $\alpha_s(0)$  & $N_f$ & $\mu_\text{g}$ (GeV) & $\kappa$ (GeV) & $m_q$ (GeV) & rms (MeV) & $\overline{\delta_j M}$ (MeV) &
$N_\text{exp}$ &
 $N_{\max}=L_{\max}$ \\ 
 \colrule
 
  \multirow{4}{*}{$c\bar c$} &   \multirow{4}{*}{0.6}  & \multirow{4}{*}{4} & \multirow{4}{*}{0.02} & 0.985 & 1.570 & 41   & 15 & 
\multirow{4}{*}{8 states} & 8 \\ 
  &    &                              &   & 0.979 & 1.587 & 32  & 21  & &16 \\ 
  &   &                              &  & 0.972 & 1.596 & 31 & 17 & & 24 \\ 
  &  &                              &  & 0.966 & 1.603 & 31   & 17  & & 32 \\ 
 %
 \colrule 
 \multirow{4}{*}{$b\bar b$} &  \multirow{4}{*}{0.6}  & \multirow{4}{*}{5} & \multirow{4}{*}{0.02} & 1.387 & 4.894 & 48  & 6  &
\multirow{4}{*}{14 states} & 8 \\ 
   & &                                  & & 1.392 & 4.899 & 41  &  6  & & 16 \\  
  & &                                   & & 1.390 & 4.901 & 39  & 7  & & 24 \\  
 
  & &                                   & & 1.389 & 4.902 & 38  & 8  & & 32 \\ 
 
    
 \botrule  
 \end{tabular}
 \end{table}

\subsection{Decay Constants}\label{sec:decay}

\begin{figure}
\centering 
\includegraphics[width=0.65\textwidth]{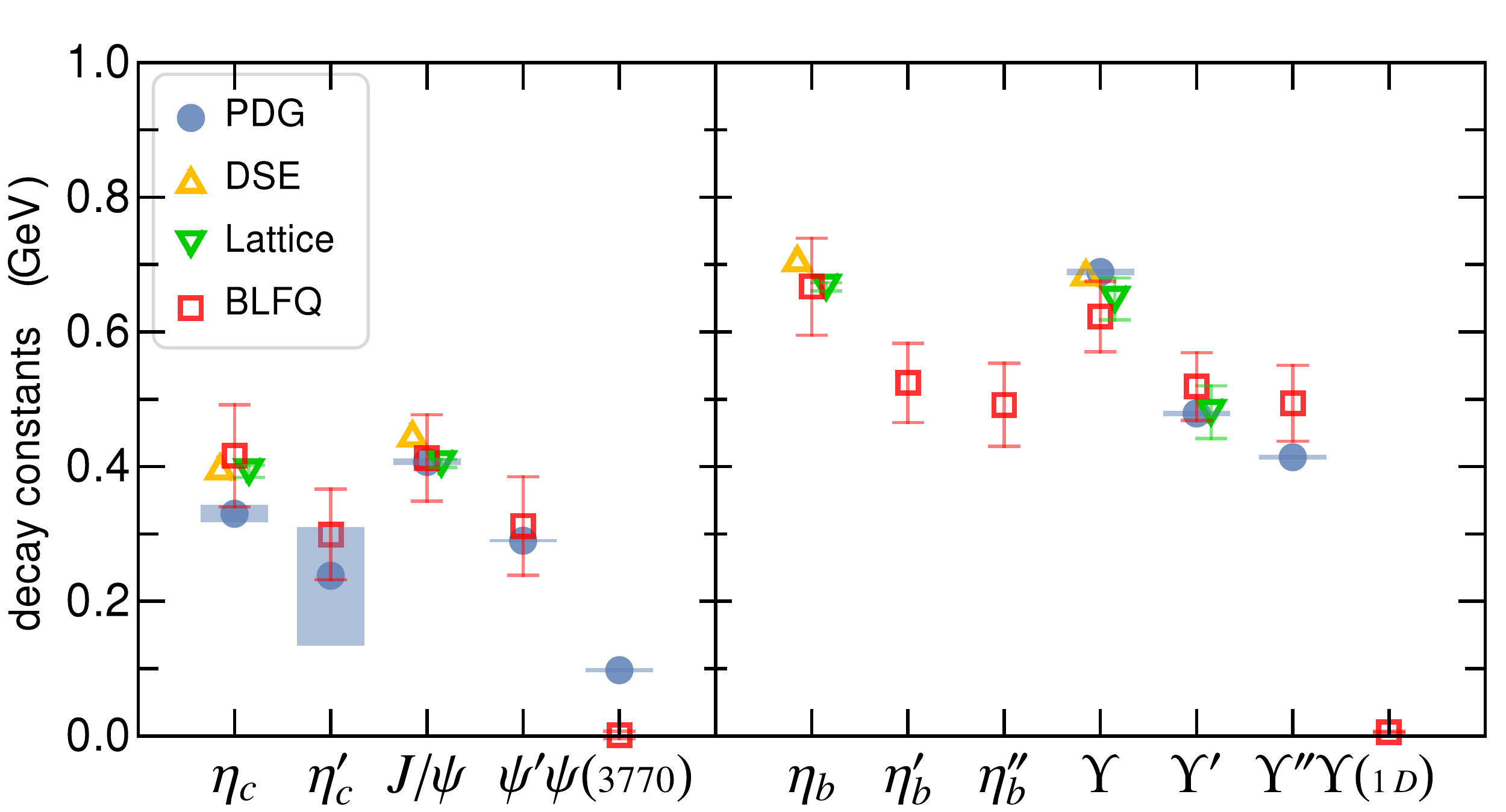}
\caption{The decay constants for vector and pseudo-scalar charmonia and bottomonia. The results are obtained with $N_{\max}=L_{\max}=8$
for charmonium and
$N_{\max}=L_{\max}=32$ for bottomonium, corresponding to UV cutoffs $\Lambda_\textsc{uv} \triangleq \kappa \sqrt{N_{\max}} \approx 1.7
m_q$, where $m_q$ is the heavy quark mass. The widths of the ``error bars'' are taken to be $\Delta f_{c\bar c} = \big|f_{c\bar
c}(N_{\max}=8) - f_{c\bar c}(N_{\max}=16)\big|$ for charmonium and $\Delta f_{b\bar b} = 2\big|f_{b\bar b}(N_{\max}=32) - f_{b\bar
b}(N_{\max}=24)\big|$ for bottomonium. They are used to indicate the sensitivity with respect to the basis truncation, rather than the full
error estimates. Results from PDG \cite{Agashe:2014kda}, Lattice \cite{Davies:2010ip,McNeile:2012qf,Donald:2012ga,Colquhoun:2014ica} and
Dyson-Schwinger equations (DSE) \cite{Blank:2011ha} are provided for comparison. 
}
\label{fig:decay}
\end{figure}

Decay constants are defined as the local vacuum-to-hadron matrix elements: 
\begin{align}
 \langle 0 | \overline \psi(0) \gamma^+\gamma_5 \psi(0) | P(p) \rangle =\,& \imag p^+ f_P,  \\
 \langle 0 | \overline \psi(0) \gamma^+ \psi(0) | V(p, \lambda) \rangle =\,& e^+_\lambda M_V f_V.
 \end{align}
Here only the ``good'' currents (the ``+'' component) are used. The corresponding LFWF representation reads \cite{Lepage:1980fj}, 
\begin{equation}\label{eqn:dc}
\frac{f_{P,V}}{2 \sqrt{2N_c}} =  \int_0^1 \frac{\dd x}{2\sqrt{x(1-x)}} \int\frac{\dd^2k_\perp}{(2\pi)^3}
\psi^{(\lambda=0)}_{\uparrow\downarrow\mp\downarrow\uparrow}(x, \vec k_\perp).
\end{equation}
For this calculation, we choose $N_{\max}=8$ for charmonium and $N_{\max}=32$ for bottomonium, roughly corresponding to $\Lambda_\textsc{uv}
\triangleq \kappa \sqrt{N_{\max}} \approx 1.7 m_q$, where $\Lambda_\textsc{uv}$ is the UV regulator, and $m_q$ is the heavy quark mass. This
choice is motived by the competition between the needs for both a better basis resolution and a lower UV scale since our model does
not incorporate radiative corrections. We also provide an indicator for sensitivity by altering the basis truncation parameter $N_{\max}$.
The resulting charmonium and bottomonium decay constants are shown in Fig.~\ref{fig:decay}, which also collects PDG values converted from
dilepton or diphoton decay widths \cite{Agashe:2014kda}, Lattice \cite{Davies:2010ip,McNeile:2012qf,Donald:2012ga,Colquhoun:2014ica} and
Dyson-Schwinger/Bethe-Salpeter equations (DSE/BSE or DSE, \cite{Blank:2011ha}; see also \cite{Ding:2015rkn}) results for comparison. Our
results fall into the ballpark of the PDG values as well as those from other approaches wherever available.

\subsection{Radii}

\begin{figure}
\centering 
\includegraphics[width=0.65\textwidth]{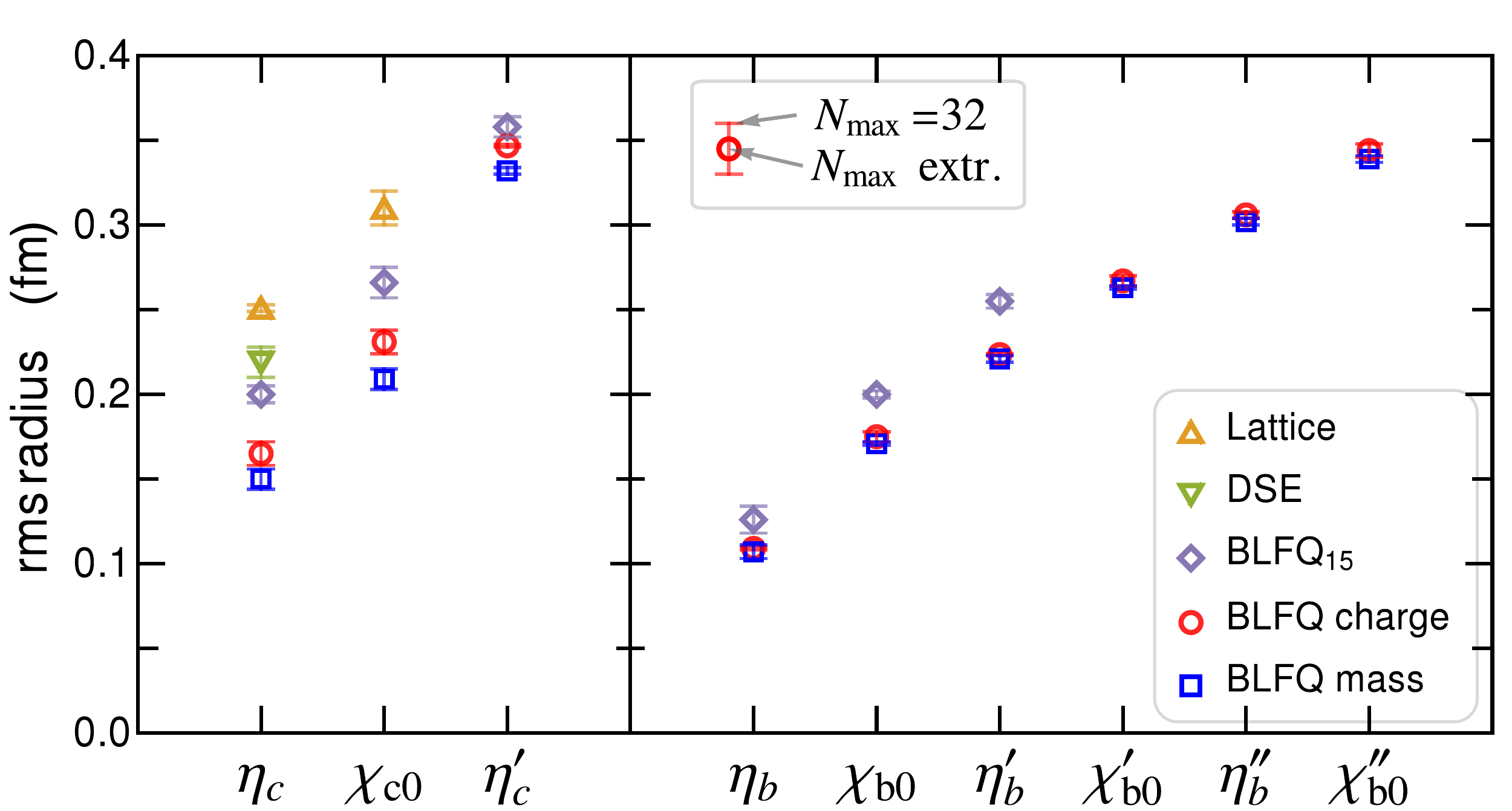}
\caption{ ``Charge'' and mass radii of (pseudo-)scalar mesons (see text). Results are obtained from extrapolating
$N_{\max}=L_{\max}=8,16,24,32$ values. The numerical uncertainty is quoted as the difference between the extrapolated result and the largest
basis result ($N_{\max}=L_{\max}=32$). Charge radii from our earlier work with fixd $\alpha_s$ (BLFQ${}_{15}$, \cite{Li:2015zda}) as well as
other approaches \cite{Dudek:2006ej, Maris:2006ea} are provided for comparison.
}
\label{fig:radii}
\end{figure}

Classically and in non-relativistic quantum mechanics, the root-mean-square charge (mass) radius is the expectation
value of the displacement operator that characterizes the charge (mass) distribution of the system. In quantum field theory, no such local
position operator is allowed and, instead, the form factors are defined as the slope of the charge (gravitational) form factor at zero
momentum transfer:
\begin{equation}\label{eqn:ff}
 \langle r^2_\mathrm{c} \rangle = -6\frac{\partial}{\partial Q^2}F_\mathrm{ch}(Q^2)\Big|_{Q\to0}, \quad 
 \langle r^2_\mathrm{m} \rangle = -6\frac{\partial}{\partial Q^2}F_\mathrm{gr}(Q^2)\Big|_{Q\to0}.
\end{equation}
Remarkably, in LFWF representation \cite{Brodsky:2000ii}, this definition exactly restores the charge (mass) distribution
interpretation \cite{Li:2016wwu}. For example, for (pseudo-)scalar mesons in the two-body approximation, 
\begin{align} 
\langle r^2_\mathrm{c}\rangle =\,& \frac{3}{2} \langle \vec b^2_\perp \rangle 
\triangleq \frac{3}{2} \sum_{s, \bar s} \int_0^1 \frac{\dd x}{4\pi} \int \dd^2r_\perp \, (1-x)^2\vec r^2_\perp \,
\widetilde\psi_{s\bar s}^*(\vec r_\perp, x)
\widetilde\psi_{s\bar s}(\vec r_\perp, x),
\label{eqn:light-front_charge_radii}\\
\langle r^2_\mathrm{m}\rangle =\,& \frac{3}{2} \langle \vec \zeta^2_\perp \rangle
\triangleq \frac{3}{2} \sum_{s, \bar s} \int_0^1 \frac{\dd
x}{4\pi} \int \dd^2r_\perp \, x(1-x)\vec r^2_\perp \,
\widetilde\psi_{s\bar s}^*(\vec r_\perp, x)
\widetilde\psi_{s\bar s}(\vec r_\perp, x).
\label{eqn:light-front_gravitational_radii}
 \end{align}
Here $\widetilde\psi$ are LFWFs in transverse coordinate space. $\vec \zeta_\perp \triangleq \sqrt{x(1-x)}\vec r_\perp$ is Brodsky and de
T\'eramond's holographic variable \cite{Brodsky:2014yha}, $\vec b_\perp \triangleq (1-x)\vec r_\perp$ is Burkardt's impact parameter
\cite{Burkardt:2000za}. This relation is also valid when higher Fock sector contributions are included if the we define $\vec \zeta_\perp$
and $\vec b_\perp$ in the $n$-body Fock sector as,
\begin{equation}
 \vec\zeta_\perp^2 \triangleq \sum_i x_i (\vec r_{i\perp} - \vec R_\perp)^2, \quad 
 \vec b_\perp^2 \triangleq \sum_i e_i (\vec r_{i\perp} - \vec R_\perp)^2,
\end{equation}
where $\vec R_\perp \triangleq \sum_i x_i \vec r_{i\perp}$ is the transverse center of the system, $e_i$ is the charge number of the $i$-th
constituent, and $\sum_i e_i \equiv Q$.

Due to charge conjugation symmetry, the charge radii of quarkonium vanishes. Here we define a fictitious charge radii by considering only
the charge of the quark. With this definition, the ``charge'' radii are the same as the mass radii in the non-relativistic limit, which
suggests that their difference is a pure relativistic effect. Fig.~\ref{fig:radii} presents the r.m.s. charge and mass radii of
scalar and pseudo-scalar mesons. In our results, the mass radii are in general smaller than the charge radii and the difference is reduced
in the
heavier system (bottomonium). Fixed $\alpha_s$ BLFQ results (BLFQ${}_{15}$, \cite{Li:2015zda}) as well as earlier results from quenched
Lattice calculation \cite{Dudek:2006ej} and DSE \cite{Maris:2006ea} are included in Fig.~\ref{fig:radii} for comparison. Our results are 
systematically smaller. From the trend with respect to basis truncation $N_{\max}=L_{\max}$, UV physics and/or higher Fock sector
contributions may be expected to produce significant corrections to our results for radii.

\section{Wave Functions, Amplitudes and Distributions}\label{sect:lfwfs}

\subsection{Light-Front Wave Functions}
Wave functions offer first-hand insight into the system. They play a central role in evaluating hadronic observables and light-cone
distributions, and are an indispensable tool for investigating exclusive processes in deep inelastic scattering \cite{Chen:2016dlk}.
Compared with the widely
used phenomenological LFWFs in the literature, our wave functions generalize the AdS/QCD wave functions and provide unified access to ground
and excited states. In particular, the spin structure is generated from the one-gluon exchange and its interplay with the confining
interaction.

In this section, we present the valence sector wave functions. Heavy quarkonium is an ideal system to explore the qualitative features of
the wave functions, as they can be compared with the familiar non-relativistic quantum mechanical wave functions. 
We show LFWFs with different polarizations and spin alignments: $\psi_{s\bar s}^\lambda(\vec k_\perp, x)$. For each spin configuration, the
\emph{orbital} angular momentum projection $m_\ell=\lambda-s_1-s_2$ is definite ($\lambda\equiv m_j$). Hence, the angular
dependence of the wave function factorizes: $\psi_{s\bar s}^\lambda(\vec k_\perp, x) = \Psi_{s\bar s}^\lambda(k_\perp, x)\exp(\imag m_\ell
\theta)$, with $\theta\equiv\arg \vec k_\perp, k_\perp \equiv |\vec k_\perp|$. 
To visualize the wave functions, we drop the phase $\exp(\imag m_\ell \theta)$, while retaining the relative sign $\exp(\imag m_\ell \pi) =
(-1)^{m_\ell}$ for negative values of $k_\perp$. Namely, we plot:
\begin{equation}
 \Psi_{s\bar s}^\lambda(k_\perp, x) \equiv \left\{
  \begin{array}{lc}
  \Psi_{s\bar s}^\lambda(k_\perp, x), & k_\perp \ge 0, \\
  \Psi_{s\bar s}^\lambda(-k_\perp, x)\times(-1)^{m_\ell}, & k_\perp<0.
 \end{array}\right. 
\end{equation}
We also define: 
$
\psi_{\uparrow\downarrow \pm \downarrow\uparrow}^\lambda(\vec k_\perp, x) \equiv \frac{1}{\sqrt{2}} \big[
\psi_{\uparrow\downarrow}^\lambda(\vec k_\perp, x) \pm \psi_{\downarrow\uparrow}^\lambda(\vec k_\perp, x) \big].
$
The full set of results is collected in supplemental materials. Here we focus on some selected results.

 \begin{figure}
 \centering 
 \subfloat[\ 
 $\psi_{\uparrow\downarrow-\downarrow\uparrow}(\vec k_\perp, x)$ ]{
 \includegraphics[width=0.37\textwidth]{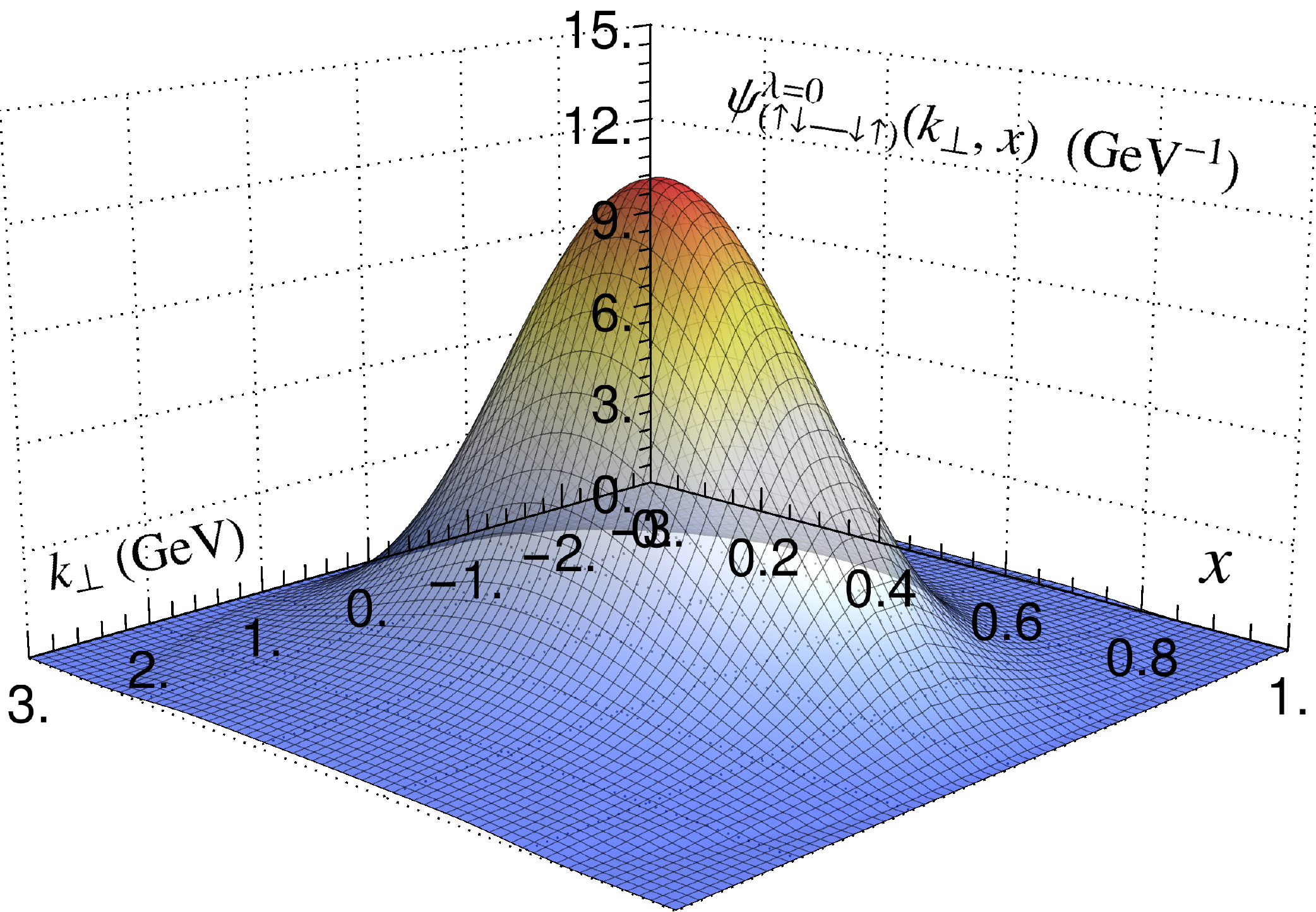}   
 \includegraphics[width=0.3\textwidth]{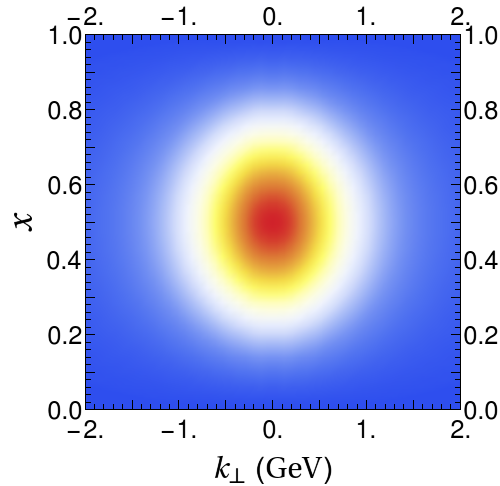}\quad
 \includegraphics[width=0.3\textwidth]{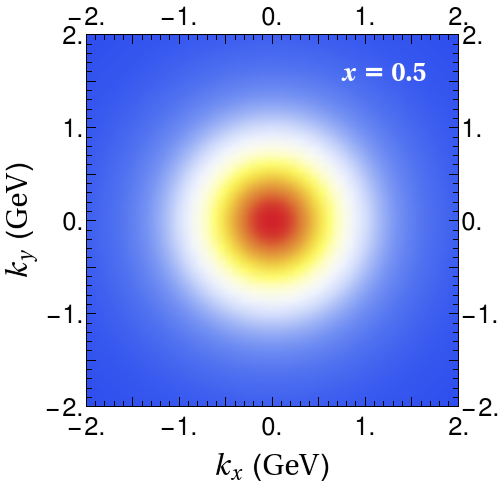}
 } 
 
 \subfloat[\ 
 $\psi_{\downarrow\downarrow}(\vec k_\perp, x)=\psi^{*}_{\uparrow\uparrow}(\vec k_\perp, x)$]{
 \includegraphics[width=0.37\textwidth]{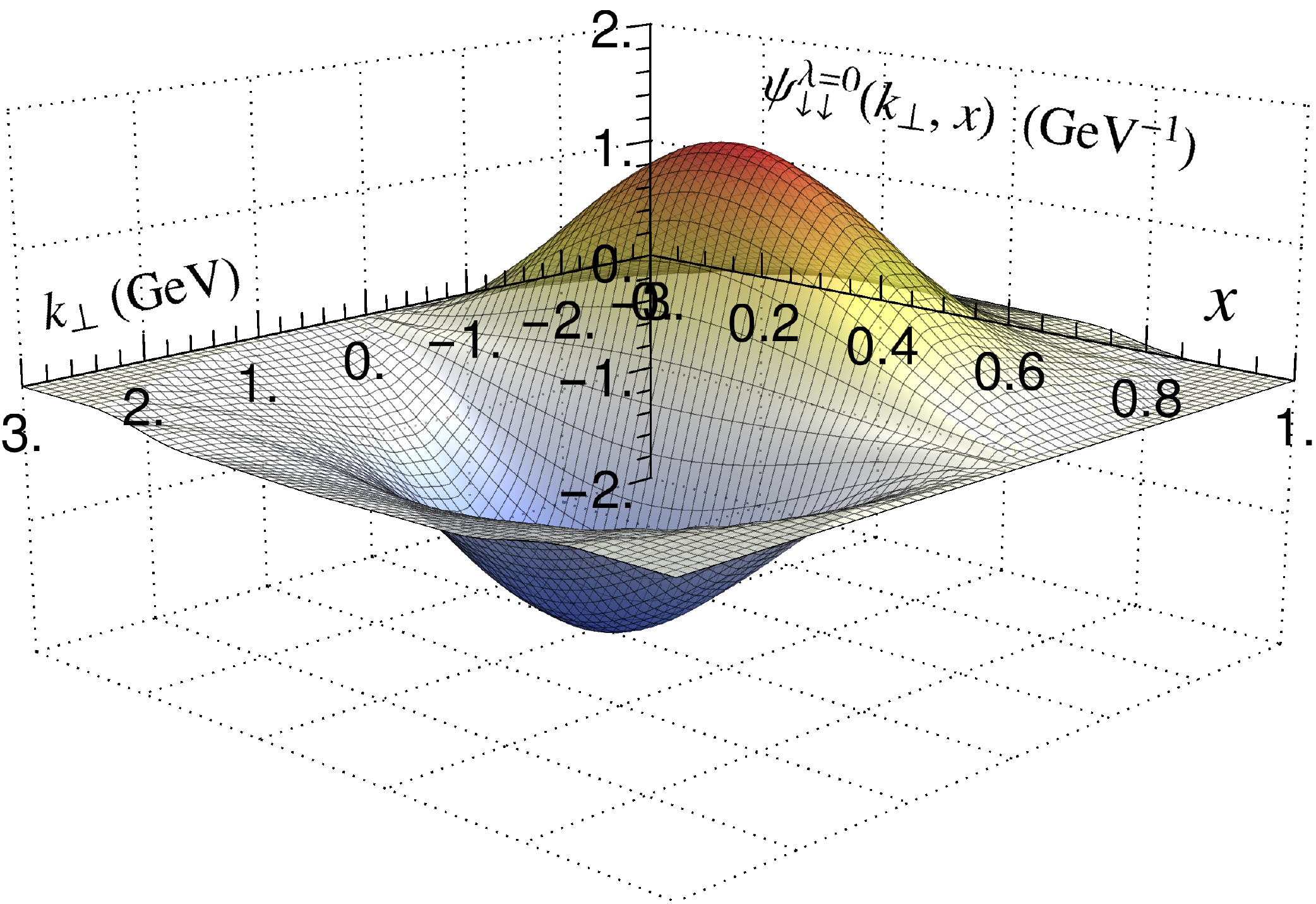} 
 \includegraphics[width=0.3\textwidth]{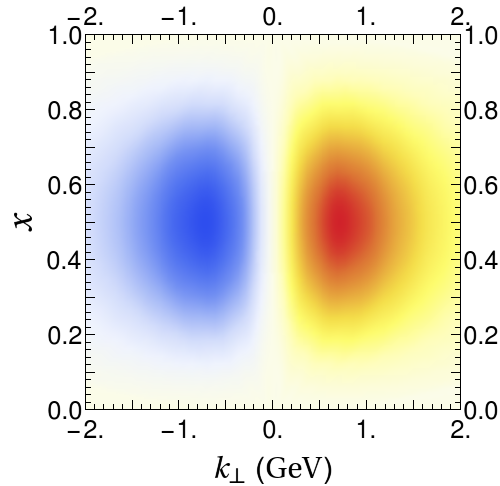}\quad
 \includegraphics[width=0.3\textwidth]{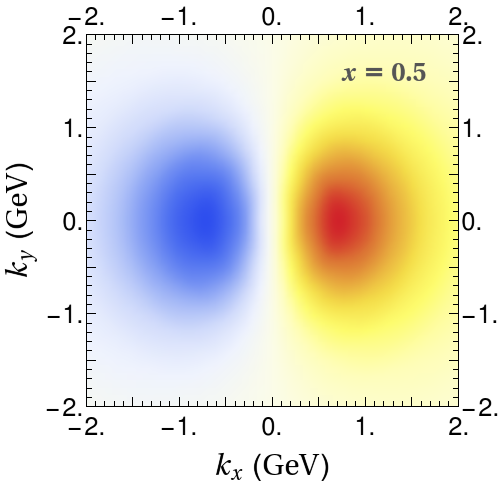}
 }
\caption{LFWFs of $\eta_c(1S)$. The left and central panels visualize LFWFs as functions of $x$ and $k_\perp$. 
The right panels show LFWFs in the transverse plane $k_x$--$k_y$ at $x=0.5$.}
\label{fig:etac1S}
\end{figure}

Figure~\ref{fig:etac1S} shows the LFWFs of the charmed ground-state pseudo scalar $\eta_c(1S)$. 
There are two independent 
components: $\psi_{\uparrow\downarrow-\downarrow\uparrow}(\vec k_\perp, x)$ and 
$\psi_{\downarrow\downarrow}(\vec k_\perp, x)=\psi^{*}_{\uparrow\uparrow}(\vec k_\perp, x)$. The number of independent components is not
\emph{a priori} the same in different relativistic approaches. One of the components is related to the non-relativistic wave functions,
whereas the other one is of purely relativistic origin and becomes negligible in the non-relativistic limit. In covariant light-front
dynamics, the extra component depends on the orientation of the quantization surface \cite{Carbonell:1998rj, Leitner:2010nx}. Its existence
ensures the rotational symmetry, albeit not exactly in our model \cite{Carbonell:1998rj}. The Lorentz structure of the pseudo scalar wave
function can be written as \cite{Carbonell:1998rj, Leitner:2010nx}, 
\begin{equation}
 \psi_{s\bar s}(\vec k_\perp, x) = \bar u_{s}(k_1)\Big[ \phi_1(k_\perp, x) \gamma_5 + \phi_2(k_\perp, x) \frac{\gamma^+\gamma_5}{P^+} \Big]
v_{\bar
s}(k_2).
 \end{equation}
where $\gamma^+ = \gamma^0+\gamma^3$. Let $n=(1,0,0,-1)$ be a null vector perpendicular to the quantization surface. 
$\gamma^+=n_\mu\gamma^\mu$, $P^+\equiv n_\mu P^\mu$, both depending on the orientation of the quantization surface.

For charmonium, the dominate component is the singlet $\psi_{\uparrow\downarrow-\downarrow\uparrow}$ and its wave function 
resembles an S-wave. In the non-relativistic limit, the longitudinal momentum fraction $x$ is reduced to: $x \to 1/2 + {k_z}/(2m_q)$.
Hence, the $x$--$k_\perp$ plots in Fig.~\ref{fig:etac1S} (central panels) are reduced to the $k_z$--$k_\perp$ density plots of the
non-relativistic wave function, i.e. a slice of the full 3D wave function, in the non-relativistic limit. To visualize the full 3D wave
function, one may rotate the density plot along the vertical axis at $k_\perp=0$, applying a phase factor $\exp(\imag m_\ell \theta)$ as
necessary\footnote{This is where the relative sign at negative $k_\perp$ is useful.}. To facilitate the visualization in 3D, we also
plot the real part of the wave functions in the transverse plane at $x=0.5$ in Fig.~\ref{fig:etac1S} (right panels).

\begin{figure}
 \centering 
 \subfloat[\ $\eta_c(1S)$]{
 \includegraphics[width=0.3\textwidth]{Figures/etac1S_mj0_singlet_N32_RunAlf_mom_style2}
}
 \subfloat[\ $\eta_c(2S)$]{
 \includegraphics[width=0.3\textwidth]{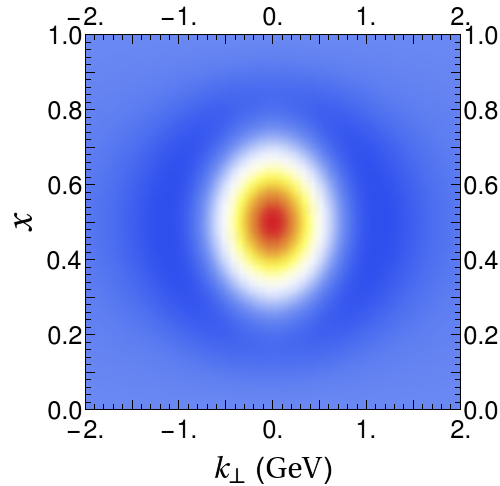}
}
 \subfloat[\ $\eta_c(3S)$]{
 \includegraphics[width=0.3\textwidth]{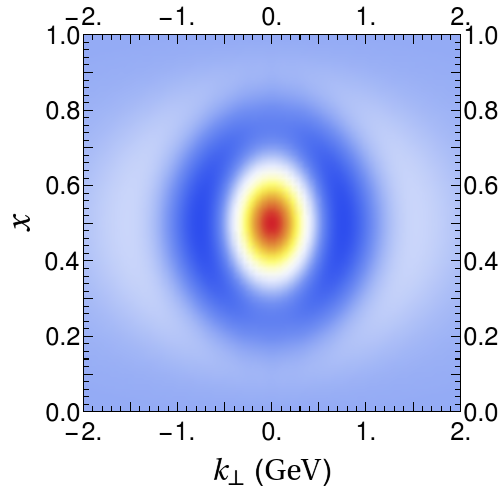}
}

 \subfloat[\ $\eta_b(1S)$]{
 \includegraphics[width=0.3\textwidth]{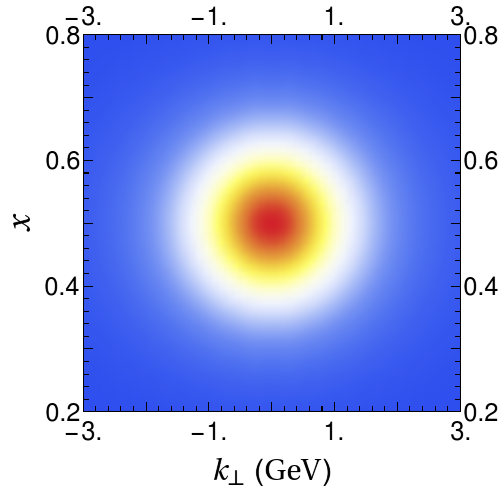}
}
 \subfloat[\ $\eta_b(2S)$]{
 \includegraphics[width=0.3\textwidth]{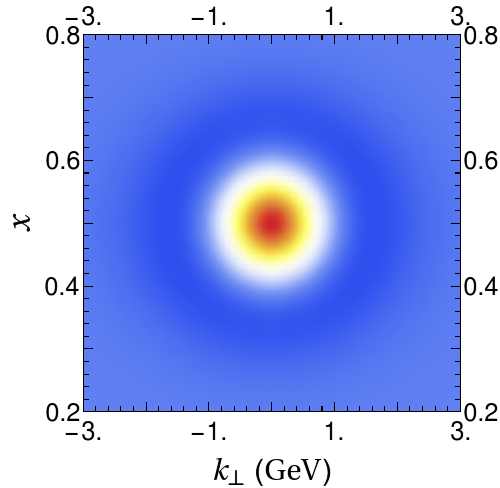}
}
 \subfloat[\ $\eta_b(3S)$]{
 \includegraphics[width=0.3\textwidth]{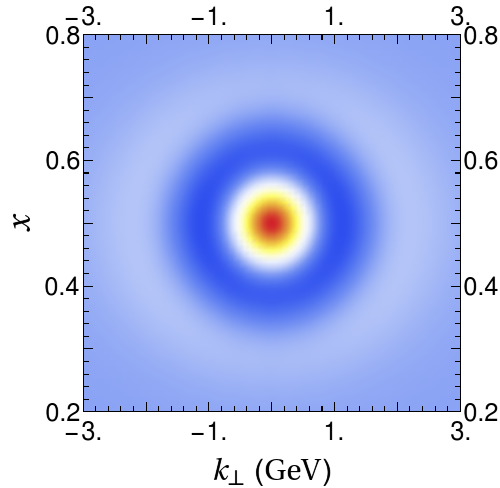}
}

\caption{Spin singlet LFWFs $\psi_{\uparrow\downarrow-\downarrow\uparrow}(\vec k_\perp, x)$ of 
charmonium (\textit{top panels}) and bottomonium (\textit{bottom panels}). }
\label{fig:etanS}
\end{figure}

Figure~\ref{fig:etanS} shows the spin singlet components of the charmed and beautified pseudo scalars $\eta_c(nS)$ and $\eta_b(nS)$. Each of
them is the dominant component in their respective systems. The 2S and 3S states show both longitudinal and transverse nodes, consistent
with the non-relativistic wave functions. Therefore, the non-relativistic picture emerges in heavy quarkonium as expected. Note that the
node structure spans a broad kinematical region [$x\sim (0.2\text{--}0.8)$] in charmonium, extending beyond the na\"ive non-relativistic
scope: $|x-\frac{1}{2}|\ll 1$. 
 
\begin{figure}
 \centering
 \includegraphics[width=0.4\textwidth]{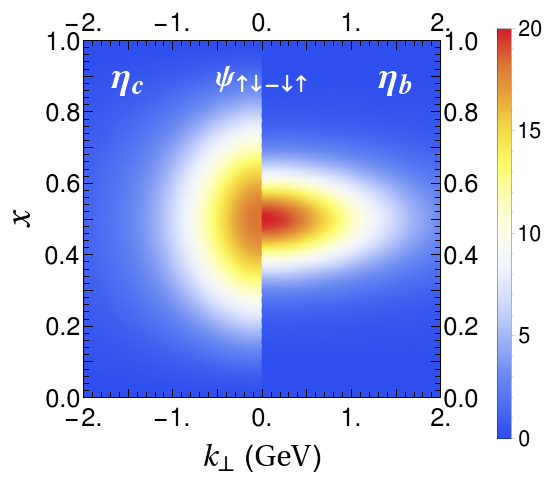}
\caption{Comparison of the spin singlet LFWFs $\psi_{\uparrow\downarrow-\downarrow\uparrow}(\vec k_\perp, x)$ between 
charmonium (\textit{left}) and bottomonium (\textit{right}). The magnitude of the wave function is in $\mathrm{GeV}^{-1}$.}
\label{fig:etac_vs_etab}
\end{figure}

 It is also interesting to compare the charmonium and bottomonium LFWFs, as shown in Fig.~\ref{fig:etac_vs_etab}.
Bottomonium is associated with a larger mass scale and is broader in the transverse momentum direction. On the 
other hand, bottomonium is more non-relativistic compared with charmonium, and hence in the longitudinal direction
its wave functions are narrower. Recall that in the non-relativistic limit, the quarkonium distribution amplitude
is a Dirac delta: $\phi(x) \propto \delta(x-\frac{1}{2})$.

\begin{figure}
 \centering 
 \subfloat[\ $J/\psi$: $\psi^{\lambda=0}_{\uparrow\downarrow+\downarrow\uparrow}$; $\ell=0, m_\ell=0$ \label{fig:VM_a}]{
 \includegraphics[width=0.3\textwidth]{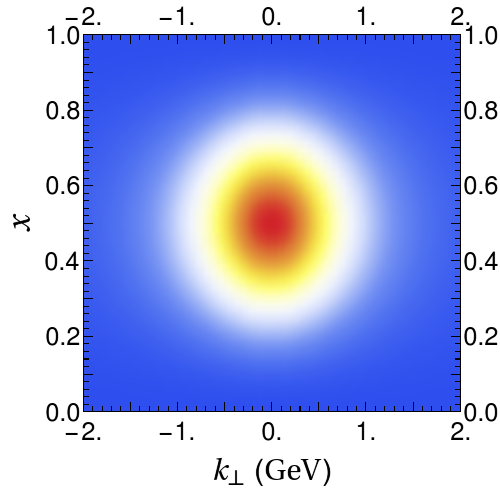}
}\quad
 \subfloat[\ $J/\psi$: $\psi^{\lambda=0}_{\downarrow\downarrow}$; $\ell=2, m_\ell=1$ \label{fig:VM_b}]{
 \includegraphics[width=0.3\textwidth]{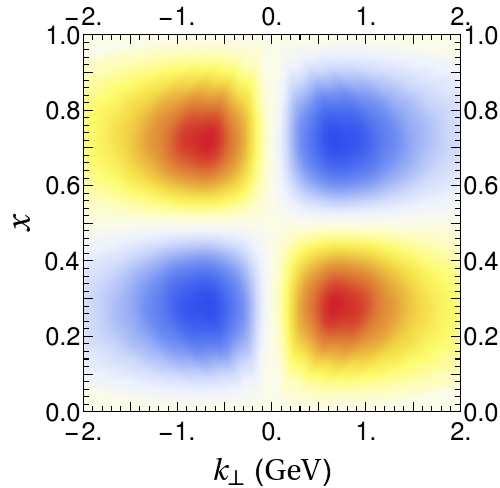}
}\quad
 \subfloat[\ $J/\psi$: $\psi^{\lambda=+1}_{\downarrow\downarrow}$; $\ell=2, m_\ell=2$ \label{fig:VM_c}]{
 \includegraphics[width=0.3\textwidth]{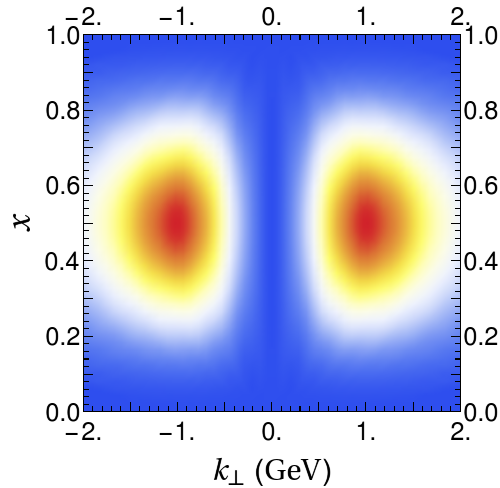}
}

 \subfloat[\ $\psi(1D)$: $\psi^{\lambda=0}_{\uparrow\downarrow+\downarrow\uparrow}$; $\ell=2, m_\ell=0$ \label{fig:VM_d}]{
 \includegraphics[width=0.3\textwidth]{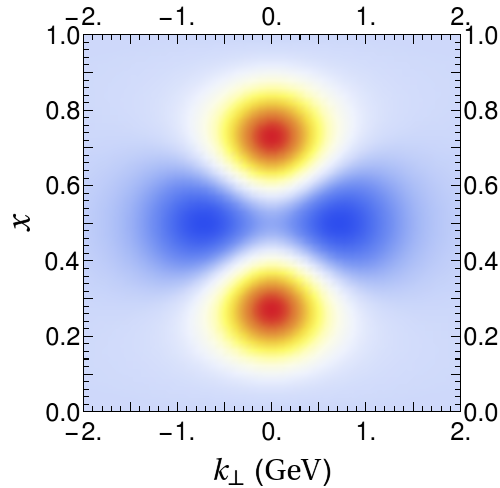}
}\quad
 \subfloat[\ $\psi(1D)$: $\psi^{\lambda=0}_{\downarrow\downarrow}$; $\ell=2, m_\ell=1$ \label{fig:VM_e}]{
 \includegraphics[width=0.3\textwidth]{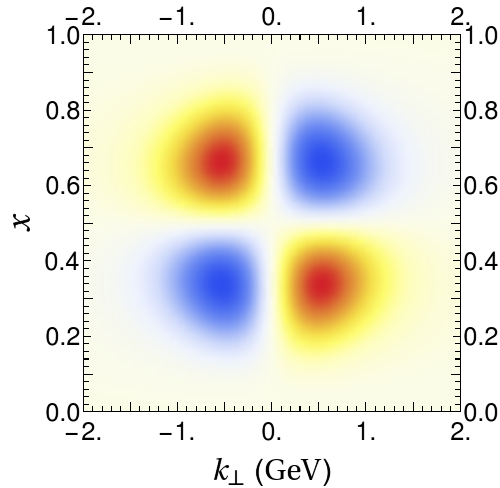}
}\quad
 \subfloat[\ $\psi(1D)$: $\psi^{\lambda=+1}_{\downarrow\downarrow}$; $\ell=2, m_\ell=2$ \label{fig:VM_f}]{
 \includegraphics[width=0.3\textwidth]{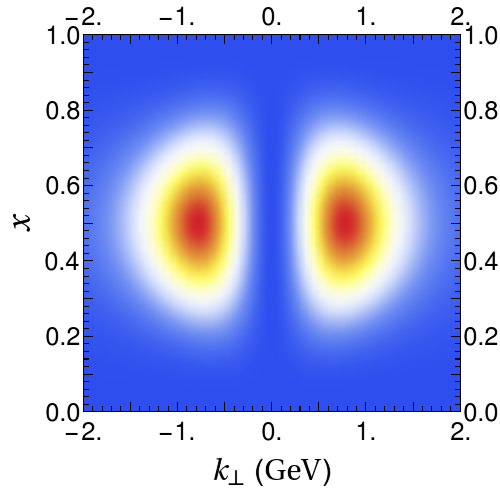}
}
\caption{Selected spin configurations of the charmed vectors $J/\psi$ (\textit{top panels}) and $\psi(1D)$ (\textit{bottom panels}).}
\label{fig:VM}
\end{figure}

\begin{figure}
 \centering
 \includegraphics[width=0.3\textwidth]{Figures/psi1D_mj1_mm_N32_RunAlf_mom_style2}
\qquad
 \includegraphics[width=0.3\textwidth]{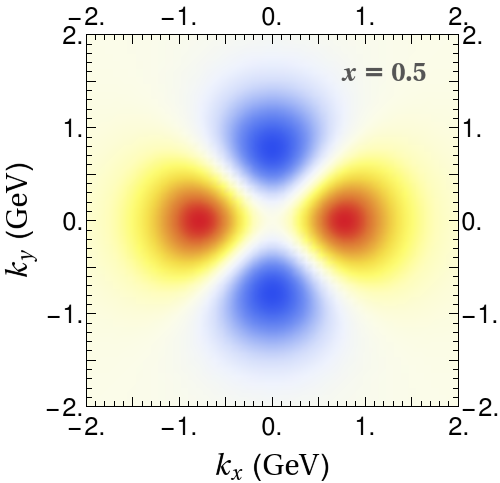}
\caption{One component of $\psi(1D)$: $\psi^{\lambda=+1}_{\downarrow\downarrow}$ in the $x$--$k_\perp$ plane (\textit{left panel}) and
in the transverse plane $k_x$--$k_y$ at $x=0.5$ (\textit{right panel}). }
\label{fig:Dwave_VM}
\end{figure}

\begin{figure}
 \centering 
 \subfloat[\ $\psi^{\lambda=0}_{\uparrow\downarrow+\downarrow\uparrow}$; $\ell=2, m_\ell=0$ \label{fig:Y2D_a}]{
 \includegraphics[width=0.3\textwidth]{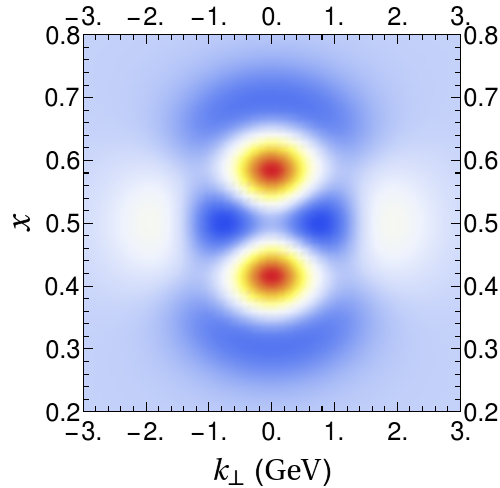}
}\quad
 \subfloat[\ $\psi^{\lambda=0}_{\downarrow\downarrow}$; $\ell=2, m_\ell=1$ \label{fig:Y2D_b}]{
 \includegraphics[width=0.3\textwidth]{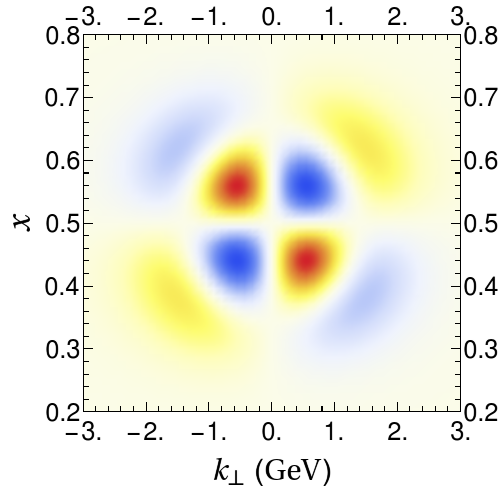}
}\quad 
\subfloat[\ $\psi^{\lambda=+1}_{\uparrow\downarrow-\downarrow\uparrow}$; $\ell=1, m_\ell=1$ \label{fig:Y2D_e}]{
 \includegraphics[width=0.3\textwidth]{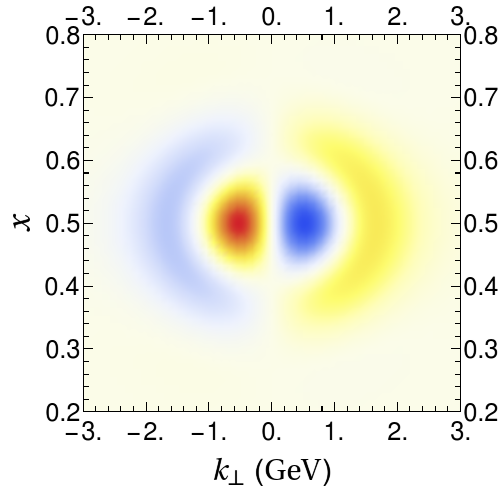}
}

 \subfloat[\ $\psi^{\lambda=+1}_{\uparrow\uparrow}$; $\ell=2, m_\ell=0$ \label{fig:Y2D_f}]{
 \includegraphics[width=0.3\textwidth]{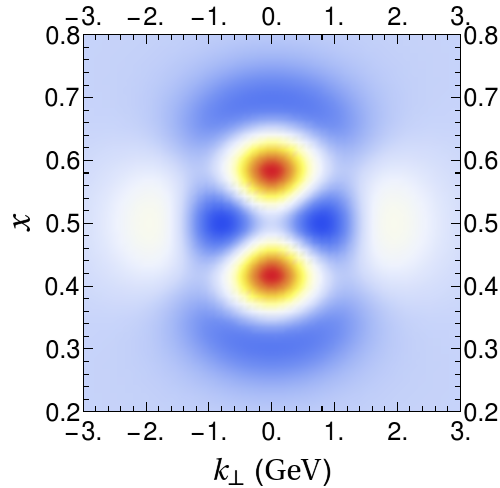}
} \quad 
 \subfloat[\ $\psi^{\lambda=+1}_{\uparrow\downarrow+\downarrow\uparrow}$; $\ell=2, m_\ell=1$ \label{fig:Y2D_d}]{
 \includegraphics[width=0.3\textwidth]{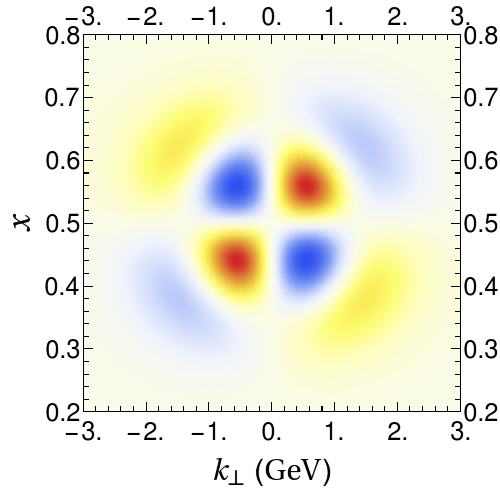}
}\quad
 \subfloat[\ $\psi^{\lambda=+1}_{\downarrow\downarrow}$; $\ell=2, m_\ell=2$ \label{fig:Y2D_c}]{
 \includegraphics[width=0.3\textwidth]{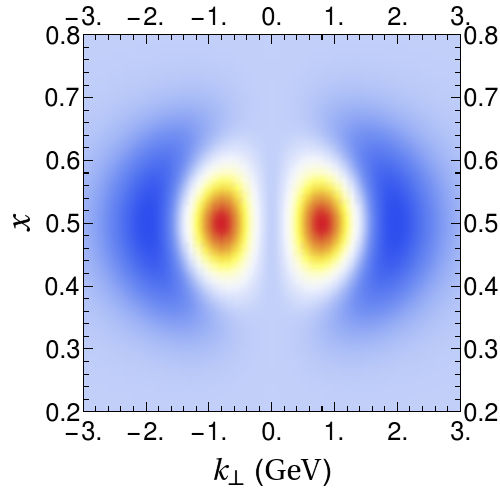}
}

\caption{The 6 independent spin components of $\Upsilon(2D)$. These wave functions show both radial and angular
excitations, in accordance with the quantum number identifications.}
\label{fig:upsilon2D}
\end{figure}

Figure~\ref{fig:VM} compares selected spin configurations of the charmed vector mesons: $J/\psi$ with its ``angular'' excitation $\psi(1D)$.
The dominant components of $J/\psi$ are $\psi^{\lambda=0}_{\uparrow\downarrow+\downarrow\uparrow}$ (Fig.~\ref{fig:VM_a}) and
$\psi^{\lambda=1}_{\uparrow\uparrow}$ (see supplemental materials) --- both are S-wave. The D-wave components (e.g.
Figs.~\ref{fig:VM_b}~\&~\ref{fig:VM_c}) are small but non-vanishing in $J/\psi$ as a result of S-D mixing. Similar sub-dominant components
due to relativity are often missing in phenomenological vector meson wave functions\footnote{Very often, the spin structure of the
phenomenological vector meson wave function is borrowed from the photon wave function, which is obtained via light-cone perturbation
theory.}, e.g., boosted Gaussian wave function \cite{Chen:2016dlk}.
The dominant components of $\psi(1D)$ are $\psi^{\lambda=0}_{\uparrow\downarrow+\downarrow\uparrow}$ (Fig.~\ref{fig:VM_d}), 
$\psi^{\lambda=0}_{\downarrow\downarrow}$ (Fig.~\ref{fig:VM_e}), 
and $\psi^{\lambda=1}_{\downarrow\downarrow}$ (Fig.~\ref{fig:VM_f}). It is evident that they resemble the non-relativistic
D-waves $Y_{20}(\hat k)$, $Y_{21}(\hat k)$ and $Y_{22}(\hat k)$, where $Y_{\ell m}(\hat k)$ are the spherical harmonics. 
This becomes more evident when LFWFs in the transverse plane ($k_x$--$k_y$) are considered (see Fig.~\ref{fig:Dwave_VM}).
Fig.~\ref{fig:upsilon2D} displays $\Upsilon(2D)$, a state consisting of both radial and angular excitations. 

\subsection{Distribution Amplitudes}\label{sec:da_and_pdf}

LFWFs provide unique access to light cone distributions by integrating out the transverse momentum \cite{Brodsky:1997de}. Among those,
the distribution amplitudes (DAs) and the parton distribution functions (PDFs) control the exclusive and inclusive processes at large
momentum transfer, respectively \cite{Lepage:1980fj}.

DAs are defined from the light-like separated gauge invariant vacuum-to-meson matrix elements \cite{Lepage:1980fj, Bodwin:2006dm}. In
light-front formalism, the leading-twist DAs within the light-cone gauge for pseudo-scalar and vector mesons\footnote{In the present work,
we focus on the longitudinal DA for vector mesons.} are \cite{Bodwin:2006dm, Braguta:2006wr, Braguta:2007fh}:
\begin{align}
 \langle 0 | \overline\psi(z)\gamma^+\gamma_5\psi(-z)|P(p)\rangle_\mu =\,& \imag p^+ f_P\int_0^1 \dd x \, e^{\imag p^+z^-(x-\half)}
\phi_P(x;
\mu)\Big|_{z^+,\vec z_\perp=0,} \\
 \langle 0 | \overline\psi(z)\gamma^+\psi(-z)|V(p,\lambda)\rangle_\mu =\,& e^+_\lambda(p) M_V f_V \int_0^1 \dd x \,
e^{\imag p^+z^-(x-\half)}
\phi_V(x;
\mu)\Big|_{z^+,\vec z_\perp=0,} \quad (\lambda=0)
\end{align}
where $f_{P,V}$ are the decay constants (see Sect.~\ref{sec:decay}). $M_{P,V}$ are the mass eigenvalues. $e^\mu_\lambda(p)$ is the
polarization vector. The non-local matrix elements as well as the DAs depend on the scale $\mu$, the renormalization scale or UV
cutoff. In these definitions, DAs are normalized to unity, viz:
\begin{equation}
 \int_0^1 \dd x \, \phi(x; \mu) = 1.
\end{equation}
In LFWF representation, DAs can be written as \cite{Lepage:1980fj},
\begin{equation}
 \frac{f_{P,V}}{2\sqrt{2N_c}} \phi_{P,V}(x; \mu) = \frac{1}{\sqrt{x(1-x)}}
\int\limits^{\mathclap{\lesssim\mu^2}} \frac{\dd^2k_\perp}{2(2\pi)^3}\psi_{\uparrow\downarrow\mp\downarrow\uparrow}^{\lambda=0}(x, \vec
k_\perp).
\end{equation}
Here $\psi_{\uparrow\downarrow\pm\downarrow\uparrow} = (\psi_{\uparrow\downarrow}\pm\psi_{\downarrow\uparrow})/\sqrt{2}$ as defined above
and the minus (plus) sign is associated with the pseudo-scalar (vector) state. 
The UV cutoff is taken as $k_\perp/\sqrt{x(1-x)}\lesssim\mu$ (see, e.g., Refs.~\cite{Lepage:1980fj,
Krautgartner:1991xz,
Zhang:1993dd}).
In the basis representation, the truncation parameter $N_{\max}$ provides a natural UV regulator $\mu\approx\kappa\sqrt{N_{\max}}$ and no
hard cutoff is needed in the integration. 

\begin{figure}
 \centering 
 \includegraphics[width=0.48\textwidth]{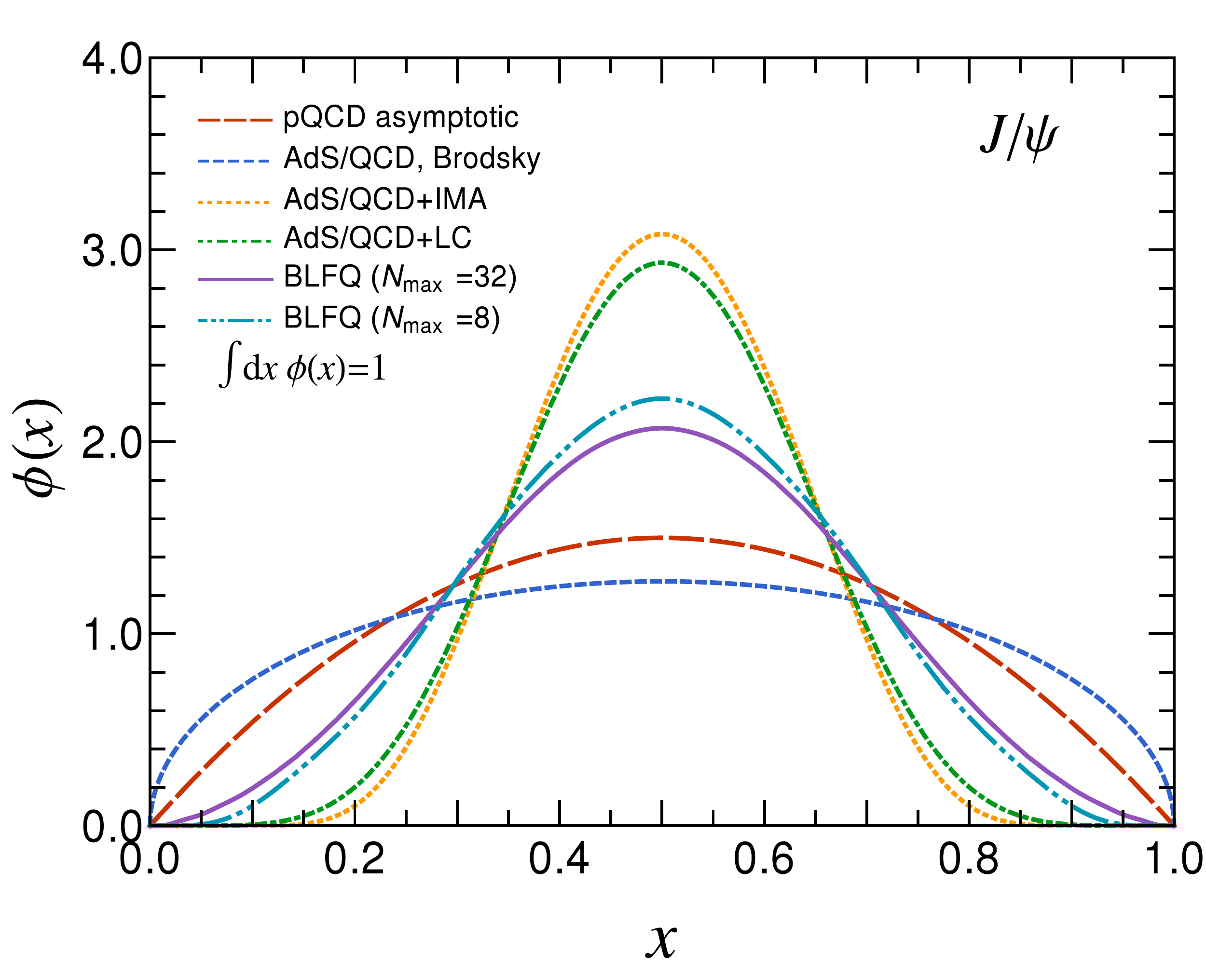}\quad
 \includegraphics[width=0.48\textwidth]{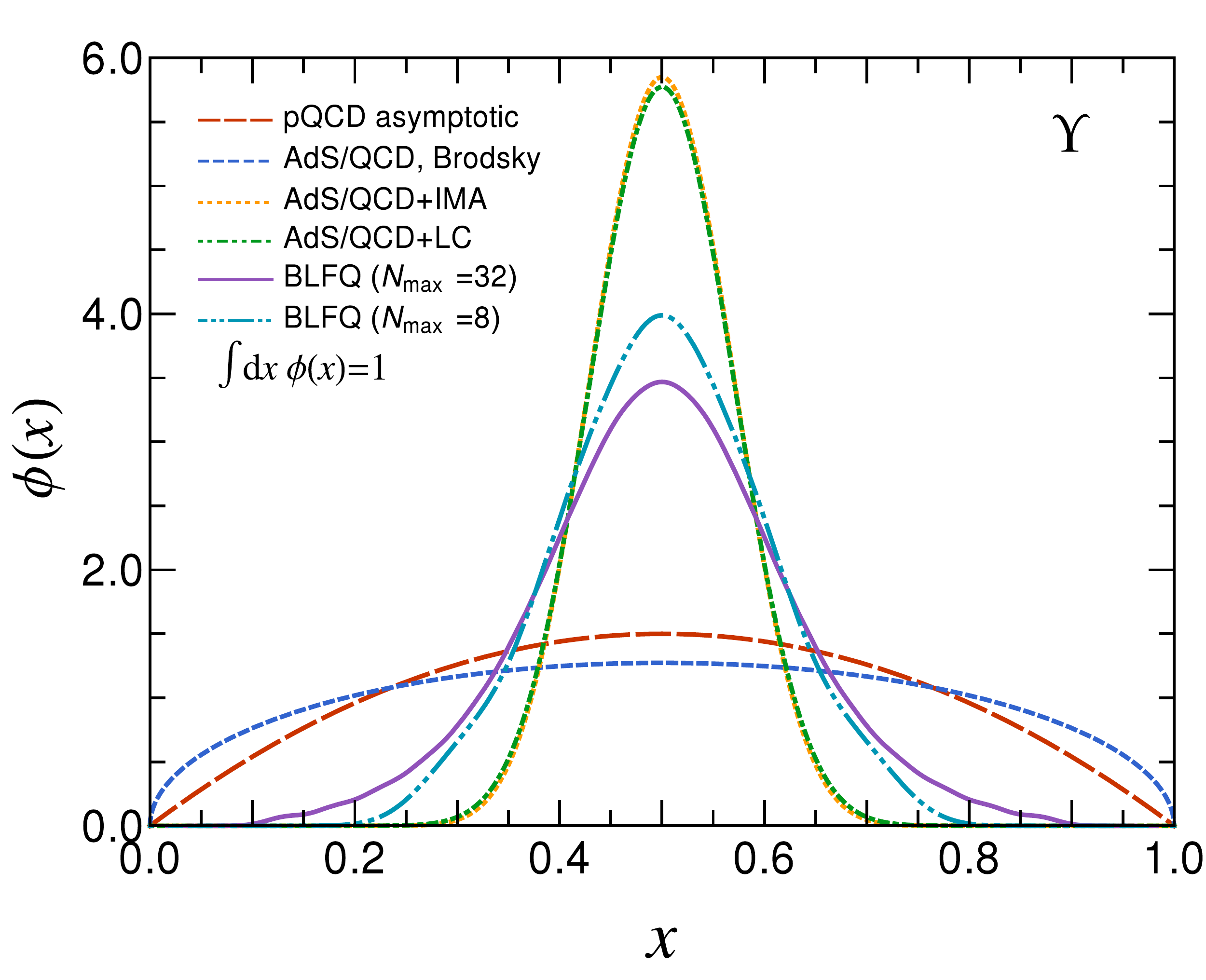}
 \caption{Comparison of the longitudinal leading-twist distribution amplitudes of $J/\psi$ (\textit{left}) and $\Upsilon$ (\textit{right}).
The pQCD asymptotic is given by $6x(1-x)$ \cite{Lepage:1980fj}. The AdS/QCD prediction of Brodsky and de T\'eramond is given by
$(8/\pi)\sqrt{x(1-x)}$ \cite{Brodsky:2014yha}. For AdS/QCD + IMA, we use parameters from Ref.~\cite{Vega:2009zb}
(cf.~\cite{Swarnkar:2015osa}) for $J/\psi$ and our parameters $\kappa, m_q$ for $\Upsilon$. AdS/QCD+LC adopts longitudinal
confinement to modify the AdS/QCD wave function, viz the basis functions. BLFQ further implements the one-gluon exchange. The BLFQ
results are with basis truncation $N_{\max}=L_{\max}=8, 32$ as indicated in the legends. The corresponding UV cutoffs are $\mu_{c\bar
c}\approx 2.8, 5.5\,\mathrm{GeV}$,
$\mu_{b\bar b}\approx 3.9, 7.9\,\mathrm{GeV}$.}
 \label{fig:DAs}
\end{figure}

\begin{figure}
 \centering 
 \includegraphics[width=0.48\textwidth]{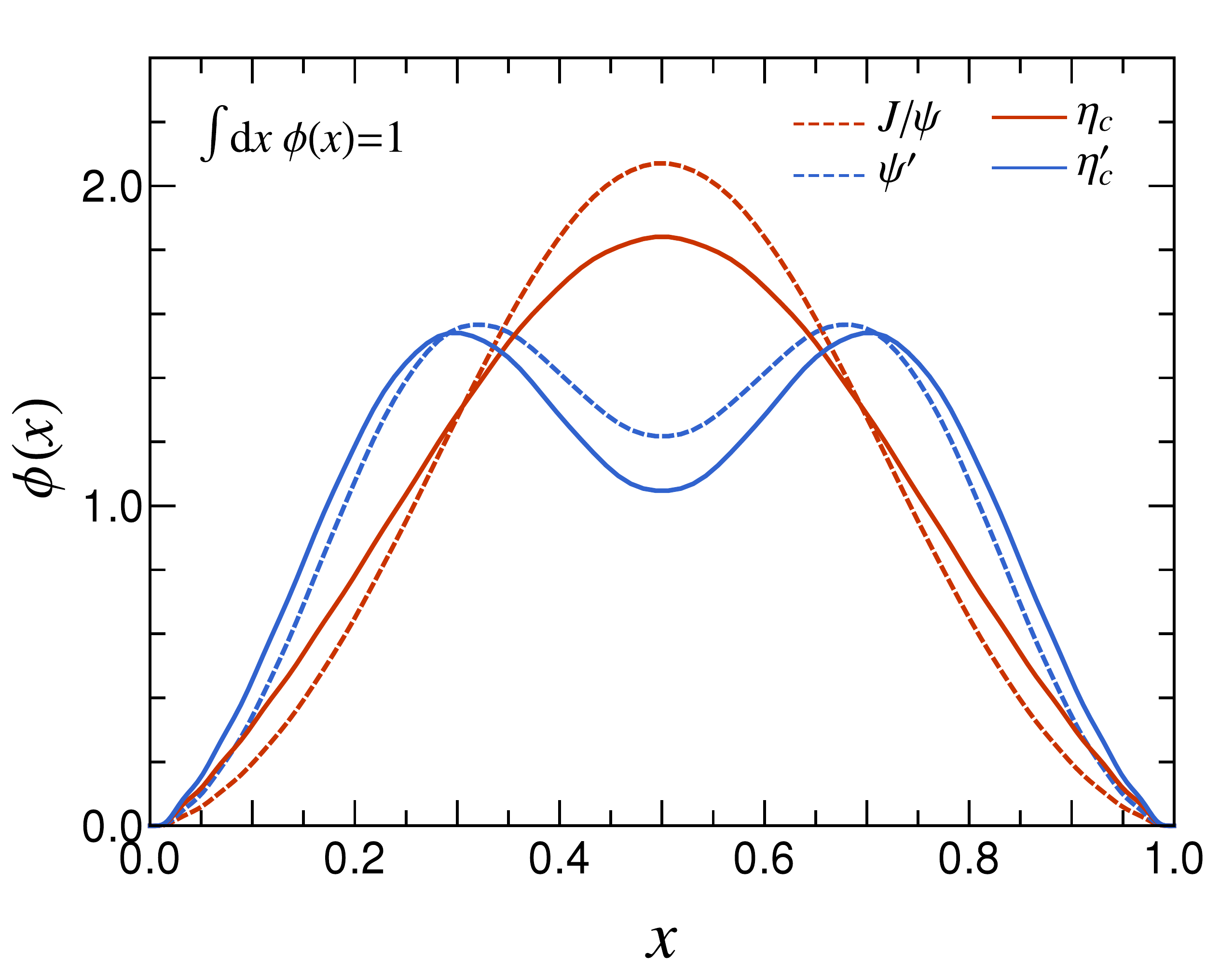}\quad
 \includegraphics[width=0.48\textwidth]{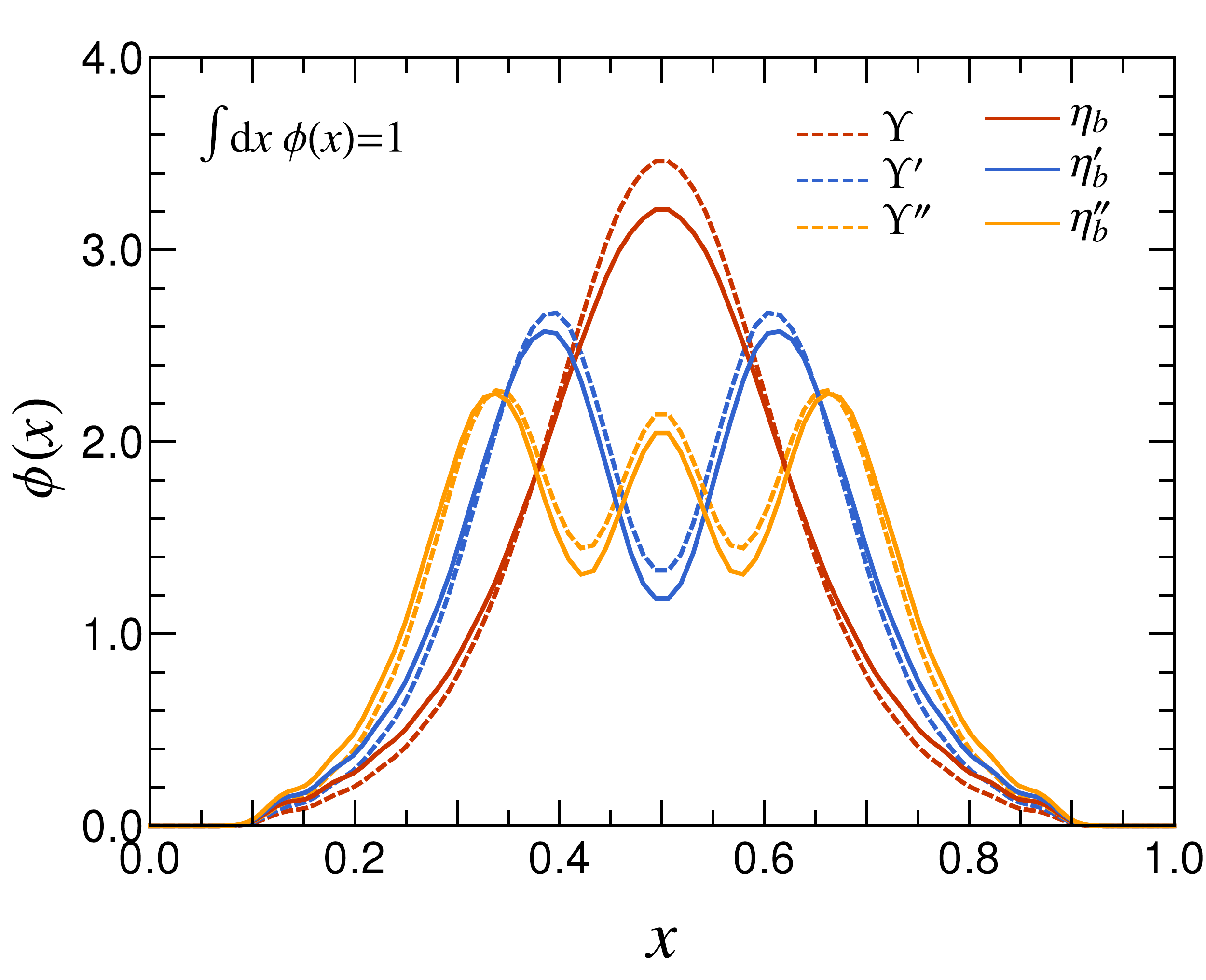}
 \caption{The leading-twist distribution amplitudes of the S-wave charmonia (\textit{left}) and S-wave bottomonia (\textit{right}) at 
$N_{\max}=L_{\max}=32$. The corresponding UV cutoffs are $\mu_{c\bar c}\approx 5.5\,\mathrm{GeV}$, $\mu_{b\bar b}\approx
7.9\,\mathrm{GeV}$.}
 \label{fig:DAs_2}
\end{figure}

Figure~\ref{fig:DAs} compares the ground-state vector meson ($J/\psi$ and $\Upsilon$) DAs with predictions from BLFQ and AdS/QCD with or
without IMA \cite{Brodsky:2014yha, Vega:2009zb, Branz:2010ub, Swarnkar:2015osa}. Calculations using pure basis functions are also
presented (AdS/QCD + LC), which turns out to be very close to AdS/QCD + IMA (cf. Fig.~\ref{fig:longitudinal_basis_functions}), but very
different from the full diagonalization (``BLFQ'') results. In fact, the BLFQ results move towards the pQCD asymptotics as the scale
increases. Obviously, the one-gluon exchange interaction plays an important role at short distance as is expected.
DAs of S-wave heavy quarkonia are shown in Fig.~\ref{fig:DAs_2}. The difference between the pseudo-scalar mesons and the
accompanying vector mesons are, again, driven by the one-gluon exchange interaction. The shape of the excited state DAs is consistent with
what has been obtained from other methods, e.g. QCD sum rule \cite{Braguta:2007tq}, wherever available. The basis functions are optimized
for long-distance
physics, i.e., confinement, and DAs are sensitive to short-distance physics. The mismatch as a finite-basis effect is clearly visible
around the endpoints in these figures.

It is useful to compute the moments in order to quantitatively compare with other approaches.
The $n$-th moment is defined as,
\begin{equation}
\langle \xi^n\rangle = \int_0^1 \dd x\, (2x-1)^n \phi(x). \qquad (\xi \equiv 2x-1)
\end{equation}
Table~\ref{tab:mom} compares the first few moments of selected heavy quarkonia states obtained from various approaches. Results from other
approaches, including non-relativistic QCD (NRQCD, \cite{Bodwin:2006dn}), QCD sum rule (QCDSR, \cite{Braguta:2006wr, Braguta:2007fh,
Braguta:2007tq}), light-front quark model (LFQM, \cite{Choi:2007ze}) and Dyson-Schwinger/Bethe-Salpeter equations (DSE,
\cite{Ding:2015rkn}), are shown for comparison. In all these approaches, moments are computed at the effective heavy quark mass scale $\mu
\simeq m_q$, with the exception of DSE at $\mu = 2\,\mathrm{GeV}$. We provide results at $\mu \approx 1.7 m_q$, corresponding to $N_{\max} =
L_{\max} = 8$ for charmonium and $N_{\max} = L_{\max} = 32$ for bottomonium. For the sake of convenience, we also provide moments at
the effective heavy quark mass scale $\mu = m_q$ (``BLFQ*'') through simple extrapolation (for charmonium) or interpolation (for
bottomonium). The $3\sigma$ (99.75\% C.L.) extrapolation or interpolation errors (prediction intervals) are included.
Our results are in reasonable agreement with various other approaches, though relativistic models, including ours, are systematically larger
than those of NRQCD. 
Results from pQCD asymptotics and AdS/QCD of Brodsky and de T\'eramond (AdS/QCD, \cite{Brodsky:2014yha}) are not particularly
applicable for heavy quarkonia at the heavy quark mass scale and are simply included for completeness.
The second moment can be used to estimate the relative velocity of the partons: $\langle v^2 \rangle \approx 3 \langle
\xi^2 \rangle$, viz
\begin{equation}
\begin{split}
 c\bar c :& \quad \langle v^2_{\eta_c} \rangle \sim 0.36, \quad \langle v^2_{\eta'_c} \rangle \sim\,  0.54; \qquad(\mu \approx 1.7 m_c) \\
 b\bar b :& \quad \langle v^2_{\eta_b} \rangle \sim 0.21, \quad \langle v^2_{\eta'_b} \rangle \sim\,  0.30,\quad \langle v^2_{\eta''_b}
\rangle \sim\,  0.36. \qquad (\mu \approx 1.6 m_b)\\
 \end{split}
 \end{equation}

\begin{table}
 \centering 
\caption{Comparison of heavy quarkonia moments from NRQCD \cite{Bodwin:2006dn}, QCD sum rule \cite{Braguta:2006wr, Braguta:2007fh,
Braguta:2007tq},
light-front quark model \cite{Choi:2007ze} and DSE \cite{Ding:2015rkn}. The DSE results are obtained at 2 {GeV}.
Results from other approaches are evaluated at quark mass scale $\mu\simeq m_q$. The BLFQ results are given at $N_{\max}=L_{\max}=8$ for
charmonium and $N_{\max}=L_{\max}=32$ for bottomonium, roughly corresponding to UV cutoffs $\mu = \kappa\sqrt{N_{\max}} \approx 1.7 m_q$.
For the convenience of comparison with other approaches, we also provide the extrapolated (ext.) or interpolated (int.) results at the
effective quark mass scale (BLFQ*). The $3\sigma$ ($\sim99.75\%$ C.L.) statistical errors (prediction intervals) are included to indicate
the quality of the extrapolation or interpolation procedure. The pQCD asymptotics \cite{Lepage:1980fj} $\langle\xi^n\rangle_\mathrm{asy} =
3/(n+1)(n+2)$ and the AdS/QCD results of Brodsky and de T\'eramond (AdS/QCD, \cite{Brodsky:2014yha}) $\langle\xi^n\rangle_\textsc{lfh} =
2(n-1)!!/(n+2)!!$, and the IMA modified AdS/QCD results (IMA, \cite{Vega:2009zb, Swarnkar:2015osa}) are also provided for comparison. 
} 
\label{tab:mom}
\begin{tabular}{cc rrrr rrrr rr}
\toprule 
 & &  {NRQCD} & {QCDSR}  & {LFQM}  & {DSE} & {BLFQ*} &  {BLFQ} & {AdS/QCD} & IMA & {pQCD} \\
\colrule   
{\multirow{4}{*}{{\large $\eta_c$}}} &  {$\langle \xi^2 \rangle$} & 0.075(11) & 0.070(7) & 0.0084{${}^{+0.004}_{-0.007}$} & 0.10 &
0.096(13) & 0.12 & 0.25 & 0.0058 & 0.20 \\
&  {$\langle \xi^4 \rangle$} & 0.010(3) & 0.012(2) & 0.017${}^{+0.001}_{-0.003}$ & 0.032 & 0.019(2) & 0.036 & 0.13 & 0.0084 & 0.086 
\\
&  {$\langle \xi^6 \rangle$} & 0.0017(7) &  0.0032(9) & 0.0047${}^{+0.0006}_{-0.0010}$ & 0.015 & 0.0036(27) & 0.014  & 0.078 & 0.0018 &
0.047
\\ 
&  {$\langle \xi^8 \rangle$} &   &    &   & 0.0059 & $-$0.0005(46) & 0.0068  & 0.055 & 0.00047 & 0.030 \\ 
& $\mu$ & $m_c$ & $m_c$  & $m_c$ & 2 GeV  & $m_c$ (ext.) & $1.7m_c$  & & & $\infty$\\
\colrule
{\multirow{4}{*}{{\large $J/\psi$}}} &  {$\langle \xi^2 \rangle$} & 0.075(11) & 0.070(7) & 0.082${}^{+0.004}_{-0.006}$ & 0.039 & 0.096(20) &
0.11 & 0.25 & 0.0058 & 0.20 \\
&  {$\langle \xi^4 \rangle$} & 0.010(3) & 0.012(2)  & 0.016${}^{+0.002}_{-0.002}$ & 0.0038 & 0.021(9) &  0.030 & 0.13 & 0.0084 & 
0.086   \\
&  {$\langle \xi^6 \rangle$} & 0.0017(7) & 0.0031(8)  & 0.0046${}^{+0.0005}_{-0.0010}$ & 7.3$\times10^{-4}$ & 0.0060(41) &  0.011 & 0.078 & 
0.0018 & 0.047  \\ 
&  {$\langle \xi^8 \rangle$} &  &   &  & 3.3$\times10^{-4}$ & 0.0015(15) &  0.0053 & 0.055 & 0.00047 & 0.030 \\ 
& $\mu$ & $m_c$ & $m_c$  & $m_c$ & 2 GeV  & $m_c$ (ext.) & $1.7m_c$ & & & $\infty$   \\
\colrule
 \multirow{4}{*}{\large $\eta_c'$} &      {$\langle \xi^2 \rangle$} & 0.22(14) & 0.18${}^{+0.005}_{-0.07}$ &  & & 0.157(9) & 0.179 \\
&      {$\langle \xi^4 \rangle$} & 0.085(110) & 0.051${}^{+0.031}_{-0.031}$ & & & 0.043(7)  &  0.059 \\
&      {$\langle \xi^6 \rangle$} & 0.039(77) & 0.017${}^{+0.016}_{-0.014}$& & &   0.013(3)  &  0.025 \\   
&      {$\langle \xi^6 \rangle$} &  &  &  &  &   0.0036(5)  &  0.012 \\   
 & $\mu$ & $m_c$ & $m_c$  & $m_c$ &  & $m_c$ (ext.)  & $1.7m_c$  &  \\
\colrule 
 \multirow{4}{*}{{\large $\eta_b$}} &        {$\langle \xi^2 \rangle$} & & & & 0.070 & 0.052(2) & 0.071 & 0.25 &  &  0.20 \\
&        {$\langle \xi^4 \rangle$} & & & & 0.015 & 0.0081(61) & 0.015   & 0.13 & &  0.086 \\
&        {$\langle \xi^6 \rangle$} & & & & 0.0042 & 0.0020(48) & 0.0051 & 0.078 &  &  0.047   \\ 
&  {$\langle \xi^8 \rangle$} &   &    &   & 0.0013 & 0.0006(31) & 0.0021  & 0.055 &  & 0.030 \\     
 & $\mu$  & $m_b$ & $m_b$ & $m_b$ & 2 GeV  & $m_b$ (int.)  & $1.6m_b$  & & & $\infty$  \\
\colrule 
 \multirow{4}{*}{{\large $\Upsilon$}} &        {$\langle \xi^2 \rangle$} & & & & 0.014 & 0.047(17) & 0.061  & 0.25 & & 0.20 \\
&        {$\langle \xi^4 \rangle$} & & & & 4.3$\times10^{-4}$ & 0.0066(73) & 0.012 & 0.13 &  &  0.086  \\
&        {$\langle \xi^6 \rangle$} & & & & $4.4\times10^{-5}$ & 0.0014(63) & 0.0036 & 0.078 & &  0.047  \\ 
&  {$\langle \xi^8 \rangle$} &   &    &   & $3.7\times10^{-6}$ & 0.0004(30) & 0.0014 & 0.055 &  &  0.030 \\     
 & $\mu$  & $m_b$ & $m_b$ & $m_b$ & 2 GeV  & $m_b$ (int.)  & $1.6m_b$ & & &  $\infty$  \\
\colrule 
 \multirow{4}{*}{{\large $\eta_b'$}} &        {$\langle \xi^2 \rangle$} & & & &  & 0.082(13) & 0.10 \\
&        {$\langle \xi^4 \rangle$} & & & &  & 0.013(15) & 0.022 \\
&        {$\langle \xi^6 \rangle$} & & & &  & 0.003(10) & 0.0068 \\ 
&  {$\langle \xi^8 \rangle$} &   &    &   &  & 0.0007(44) & 0.0027 \\     
 & $\mu$  & $m_b$ & $m_b$ & $m_b$ & 2 GeV  & $m_b$ (int.)  & $1.6m_b$  &  \\
\botrule 
\end{tabular}
\end{table}

\subsection{Parton Distributions}
The quark Parton Distribution Function (PDF) $q(x; \mu)$ is the probability of finding a collinear quark carrying momentum fraction $x$ up
to
scale $\mu$.
In the light-front formalism, it can be obtained by integrating out the transverse momentum of the squared wave function:
\begin{equation}
q(x; \mu) = \frac{1}{x(1-x)} \sum_{s, \bar s}\int\limits^{\mathclap{\lesssim\mu^2}} \frac{\dd^2k_\perp}{2(2\pi)^3} \big| \psi_{s\bar
s}(x, \vec k_\perp) \big|^2.
\end{equation}
Within the two-body approximation, the PDF and its first moment are normalized to unity [cf. Eq.~(\ref{eqn:normalization})]:
\begin{equation}
 \int_0^1 \dd x\, q(x; \mu) = 1, \quad \int_0^1  \dd x\, \big[ xq(x; \mu)+(1-x)q(x; \mu)\big] = 1.
\end{equation}

\begin{figure}
 \centering 
 \includegraphics[width=0.45\textwidth]{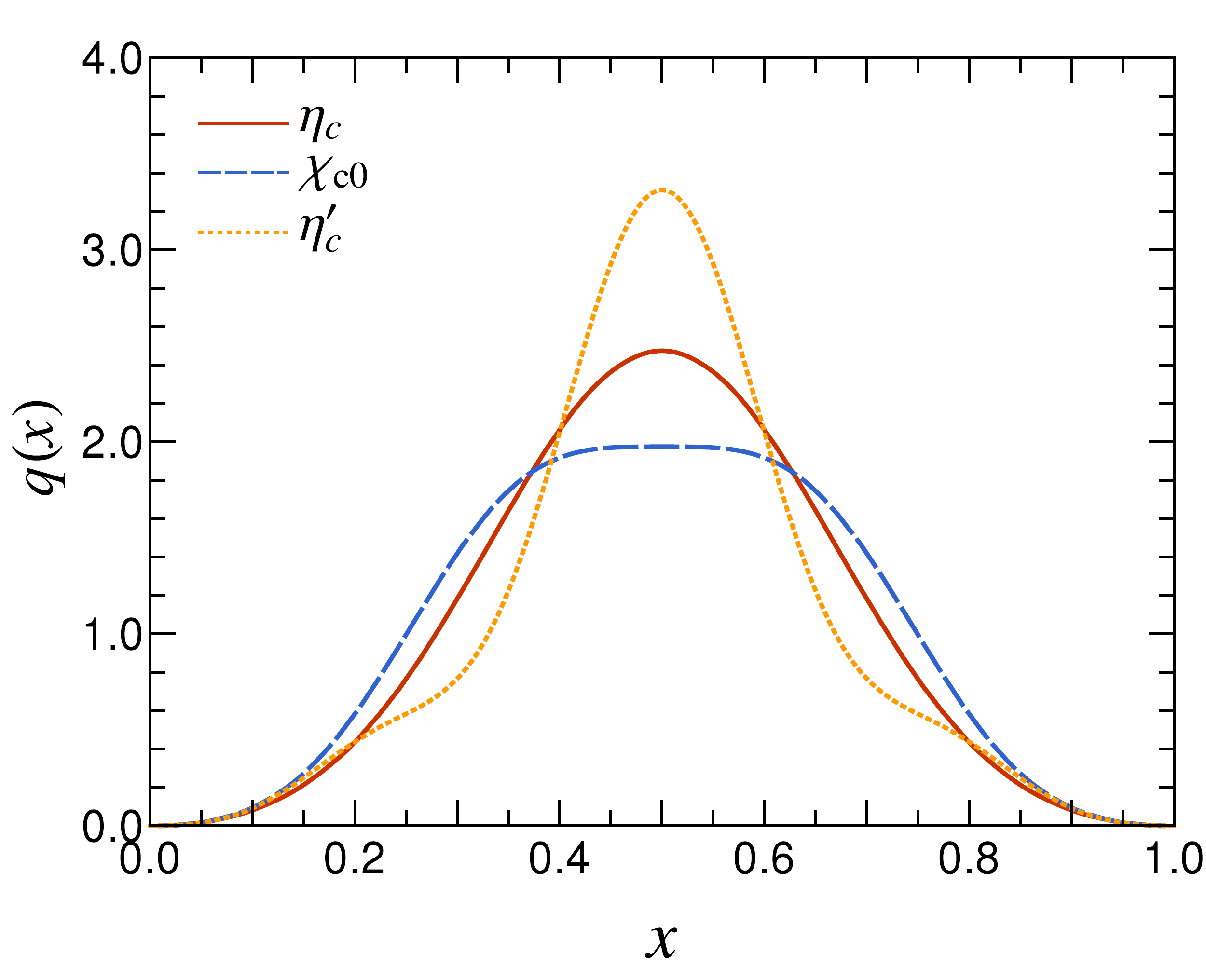} \quad 
 \includegraphics[width=0.445\textwidth]{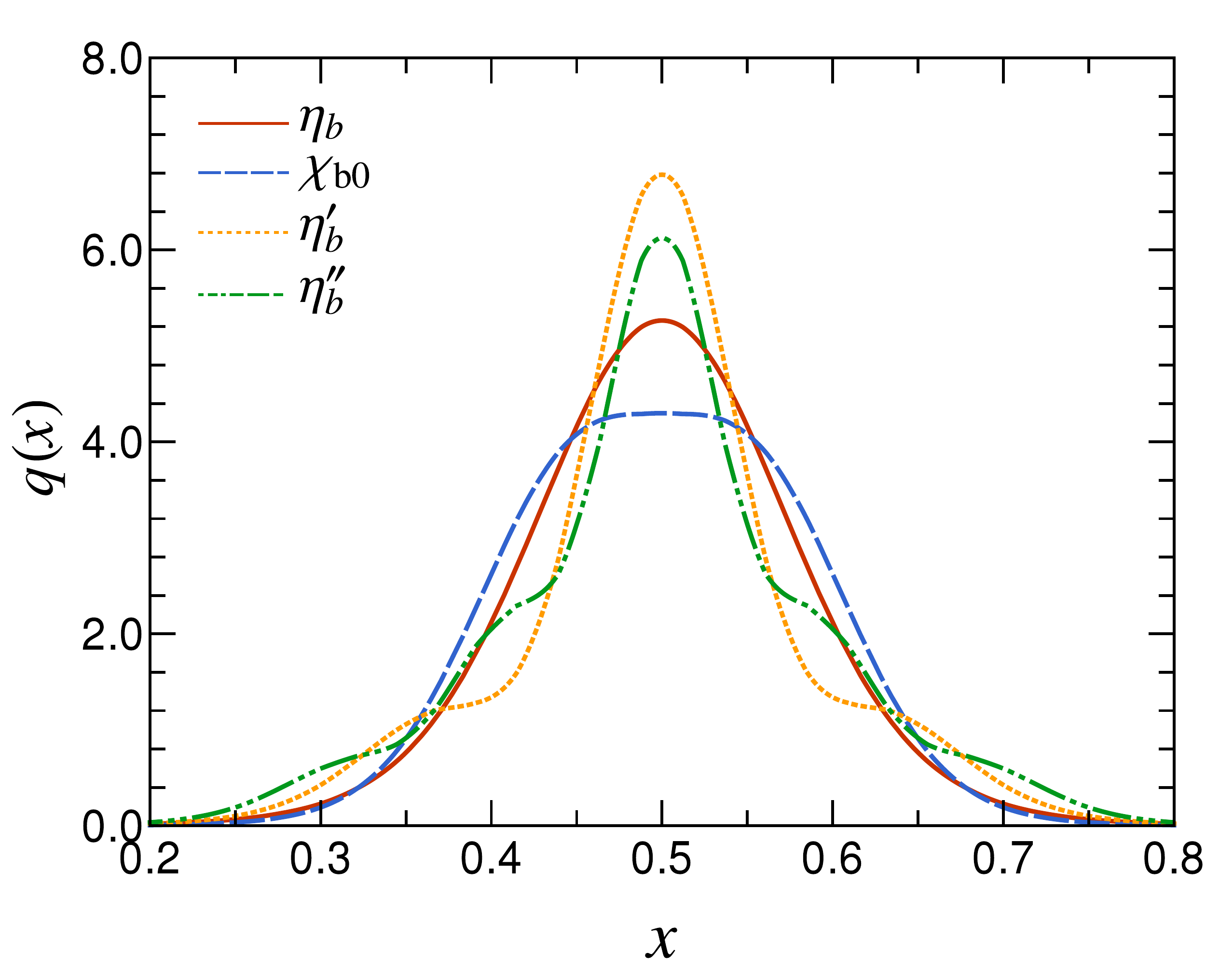}
 \caption{PDFs of (pseudo-)scalar charmonia (\textit{left}) and bottomonia (\textit{right}) at 
$N_{\max}=L_{\max}=32$. The equivalent UV cutoffs are $\mu_{c\bar c}\approx 5.5\,\mathrm{GeV}$, $\mu_{b\bar b}\approx 7.9\,\mathrm{GeV}$.}
 \label{fig:pdf}
\end{figure}

Figure~\ref{fig:pdf} shows PDFs of (pseudo-)scalar quarkonia. They exhibit distinctive features compared with DAs. In particular, there is
no dip at $x=1/2$ in excited-state PDFs, in contrast to DAs. There appear to be ripples on the downward slopes of PDFs for excited states as
may be expected from contributions of longitudinally excited basis functions.

The generalization of PDFs, known as generalized parton distributions (GPDs), unifying PDFs and form factors, provide more insights into the
system, and are directly related to experiments \cite{Burkardt:2000za, Diehl:2003ny}. Wigner distributions are more general quantities
unifying GPDs and the transverse momentum distributions. In principle, all of them are accessible through LFWFs, at least in
some kinematical regime (e.g.~\cite{Swarnkar:2015osa}). For example, in the zero skewedness limit, the impact parameter GPD
$q(x, \vec b_\perp)$ of Burkardt \cite{Burkardt:2000za} is related to the LFWFs simply by,
\begin{equation}
 q(x, \vec b_\perp) = \frac{1}{(1-x)^2}\sum_{s, \bar s} \Big| \widetilde\psi_{s\bar s}\big( \vec b_\perp/(1-x), x \big) \Big|^2.
\qquad \big( \; \vec b_\perp = (1-x) \vec r_\perp \; \big)
\end{equation}

\section{Summary and Discussions}\label{sect:summary}

We present a light-front model for quarkonium that incorporates light-front holographic QCD and the one-gluon exchange
interaction with a running coupling. We solve the model in the Hamiltonian approach with a basis function expansion. 
We obtain mass spectroscopy and the light-front wave functions. The spectroscopy agrees with the PDG data within 30--40 MeV of r.m.s. mass
deviation for states below the open flavor threshold.
The overall quality improves the previous work that employed a fixed strong coupling and a non-covariant counterterm.
The wave functions reveal rich structures, especially for excited states. Through analysis and comparison, we find
these structures are consistent with the standard non-relativistic picture. From these wave functions, we also compute
the decay constants, r.m.s. radii, distribution amplitudes, and parton distributions. Our results appear to be in 
reasonable agreement with those from other approaches wherever available.

This work is an attempt to improve light-front holographic QCD approach by adding realistic QCD interactions. In particular, we show that
while rotational symmetry is broken due to truncation, the extraction of angular momentum $j$ is feasible and reliable
\cite{Trittmann:1997xz, Brodsky:2006ez}. The Hamiltonian formalism and the basis function approach enable us to access a
wide range of states, including radial and angular excited states extending over all known excited states and beyond. The obtained
light-front wave functions allow us to directly compute hadronic distributions such as distribution amplitudes as well as hadronic
observables. It should be emphasized that these attractive features are not limited to the present effective model---they are the shared
advantages within the light-front Hamiltonian formalism \cite{Bakker:2013cea}. 

We did not include self-energy in solving the heavy quarkonia. However, radiative corrections may become important in evaluating some
observables as we employ more realistic field-theory dynamics. The calculation of the decay constants illustrates this particular challenge.
As we move to the light sector, the consistent inclusion of self-energies and renormalization issues may become more acute if one wants
to address additional phenomena within QCD such as chiral symmetry breaking. Nevertheless, we believe the present work may serve
as a substantial step for developing an elaborate light-front model for hadrons as relativistic bound states. 

While the advantages of the basis function expansion is obvious, it nevertheless requires more investigation. The IR and UV scales are tied
to the basis truncation parameter $N_{\max}$ and $L_{\max}$. Compared to the wave-equation approach, the UV asymptotics is not easy to
analyze. We typically rely on extrapolation of the basis parameters as developed in ab initio nuclear structure calculations
\cite{Coon:2012ab}. In BLFQ, the basis extrapolation requires further study. One investigation was conducted in the context of
strong coupling light-front QED and the authors found robust basis extrapolations that are consistent with the
wave-equation approach \cite{Wiecki:2014ola}. The coupling ($\alpha=0.3$), the transverse basis as well as the one-photon exchange kernel
used in Ref.~\cite{Wiecki:2014ola} are very similar to the present model. 

Future developments should focus on the inclusion of higher Fock sectors and the non-perturbative renormalization (see
Ref.~\cite{Hiller:2016itl} for a recent review). In the top-down approach, a systematic non-perturbative renormalization scheme should be
developed and non-pertubative dynamics has to be addressed using efficient numerical methods. Notable examples include the full basis
light-front quantization (BLFQ, \cite{Vary:2009gt}), the renormalization group procedure for effective particles (RGPEP,
\cite{Glazek:2012qj, Gomez-Rocha:2015esa}), the Fock sector dependent renormalization (FSDR, \cite{Karmanov:2008br, Li:2015iaw}), and the
light-front coupled cluster method (LFCC, \cite{Chabysheva:2011ed}). In the bottom-up approach, one is motivated to design appropriate
kernels that incorporate important physics while preserving the symmetries. Notable physics goals for hadrons include the radiative
corrections, asymptotic freedom and the dynamical chiral symmetry breaking. Incorporating the running coupling is the first step. In both
approaches, the current model may serve as a first approximation. See also Refs.~\cite{Ji:2012ux, Chang:2013pq,
Ji:2013dva, Karmanov:2005nv, Leitao:2017} for some recent works bridging other approaches with the light-front approach. 

The applicability of the current model is not restricted to heavy quarkonium. 
Extensions to other meson and baryon systems, in principle, are straightforward, although 
new issues have to be addressed in each of these systems.

\section*{Acknowledgements}

We wish to thank X.~Zhao, G.~Chen, M.-j.~Li, S.~Leit\~ao, S.J.~Brodsky, G.~de~T\'eramond, E.~Swanson, J.R.~Spence, J.R.~Hiller,
S.S.~Chabysheva, S.D.~G\l{}azek, A.~Trawi\'nski and M.~Gomez-Rocha for valuable discussions. 
One of us (Y.L.) also wants to thank the hospitality of the High Energy Nuclear Theory Group at the Institute of Modern Physics, Chinese
Academy of Sciences, Lanzhou, China, where part of the work is being completed. 

This work was supported in part by the Department of Energy under Grant Nos.~DE-FG02-87ER40371 and DESC0008485 (SciDAC-3/NUCLEI).
Computational resources were provided by the National Energy Research Supercomputer Center (NERSC), which is supported by the Office of
Science of the U.S. Department of Energy under Contract No.~DE-AC02-05CH11231.

\appendix

\section{Light-Front Coordinates}
We adopt natural units throughout the article: $\hbar = c = 1$. We roughly follow the convention of Ref.~\cite{Wiecki:2014ola}.
The light-front coordinates are defined as $x = (x^-, x^+, x^1, x^2) \equiv (x^-, x^+, \vec x_\perp)$, where $x^\pm = x^0 \pm x^3$. The
inner product of two 4-vectors is defined as: $a\cdot b = \frac{1}{2} a^- b^+ + \frac{1}{2} a^+ b^- - \vec a_\perp \cdot \vec b_\perp$. It
should be noted that the determinant of the metric tensor is $\det g = -(1/4)$.

The Lorentz invariant phase space measure is 
\begin{equation}
 \int \frac{\dd^4 p}{(2\pi)^4}\vartheta(p^0) 2\pi\delta(p^2-m^2) = 
 \int \frac{\dd^3 p}{(2\pi)^32p^0}\vartheta(p^0) =
 \int \frac{\dd^2 p_\perp \dd p^+}{(2\pi)^32p^+}\vartheta(p^+), \quad (m^2\ge0)
\end{equation}
where $\vartheta(z)$ is the unit step function. The one-particle state is normalized as: $\langle p, j, m_j |  p', j', m_j'\rangle =
2p^+\vartheta(p^+)(2\pi)^3\delta^3(p-p')\delta_{jj'}\delta_{m_j,m_j'}$, where the Dirac delta is defined as $\delta^3(p) \equiv
\delta(p^+)\delta^2(\vec p_\perp)$.

\section{Few-Body Kinematics}

We define boost-invariant momenta from the single-particle momenta $\{p^+_i, \vec p_{i\perp}\}$ as,
\begin{equation}
 x_i = p^+_i/P^+, \quad \vec k_{i\perp} = \vec p_{i\perp} - x_i \vec P_\perp. \qquad (P^+ = \sum_i p^+_i, \quad \vec P_\perp = \sum_i \vec
p_{i\perp})
\end{equation}
$x_i$ are the longitudinal light-front momentum fractions; and $k_{i\perp}$ are the transverse relative momenta. They satisfy:
\begin{equation}
 \sum_i x_i = 1, \quad \sum_i \vec k_{i\perp} = 0.
\end{equation}
The $n$-body phase space integration measure factorizes:
\begin{equation}
  \prod_{i} \int \frac{\dd^2 p_{i\perp} \dd p^+_i}{(2\pi)^32p^+_i}\vartheta(p^+_i) =   \int \frac{\dd^2 P_\perp \dd
P^+}{(2\pi)^32P^+}\vartheta(P^+) \prod_{i} \int_0^1\frac{\dd x_i}{2x_i}\int \frac{\dd^2 k_{i\perp}}{(2\pi)^3}
\times 2(2\pi)^3\delta\Big(\sum_i x_i-1\Big)\delta^2\Big(\sum_i\vec k_{i\perp}\Big).
\end{equation}
The invariant mass squared of the $n$-body Fock state is:
\begin{equation}
 s \equiv (p_1 + p_2 + \cdots p_n)^2 = \sum_i \frac{\vec k_{i\perp}^2+m^2_i}{x_i}. \qquad (p^2_i = m^2_i)
\end{equation}

\section{Spinors}

The $u$, $v$ spinors are defined as,
\begin{equation}
u_s(p) = \frac{1}{2\sqrt{p^+}}(\slashed{p} + m ) \gamma^+ \chi_s, \qquad 
v_s(p) = \frac{1}{2\sqrt{p^+}}(\slashed{p} - m ) \gamma^+ \chi_{-s},
\end{equation}
where $\chi_{+} = (1,0,0,0)^\transpose, \chi_{-} = (0,1,0,0)^\transpose$; $\gamma^\pm = \gamma^0 \pm \gamma^3$;  
$s=\pm$ is the light-front helicity. 
The $u$, $v$ spinors defined above are polarized in the $z$-direction (or longitudinal direction):
\begin{equation}
S_z u_\pm(p^+, \vec{p}_\perp=0) = \pm \half u_\pm (p^+, \vec{p}_\perp=0), \quad 
S_z v_\pm(p^+, \vec{p}_\perp=0) = \mp \half v_\pm (p^+, \vec{p}_\perp=0), \\
\end{equation}
$S_z\equiv \frac{\imag}{2}\gamma^1\gamma^2$ and follow the standard orthonormality 
\begin{equation}
 \bar{u}_s(p) u_{s'}(p) = 2m\delta_{s s'}, \quad
 \bar{v}_s(p) v_{s'}(p) = -2m\delta_{s s'},\quad
 \bar{u}_s(p) v_{s'}(p) = \bar{v}_s(p) u_{s'}(p) = 0,
\end{equation}
and completeness
\begin{equation}
 \sum_{s=\pm} u_s(p) \bar{u}_s(p) = \slashed{p} + m,\qquad
 \sum_{s=\pm} v_s(p) \bar{v}_s(p) = \slashed{p} - m.
\end{equation}
Here are some useful identities:
\begin{equation}
 \bar u_{s'}(p')\gamma^+u_s(p) = 2\sqrt{p^+p'^+}\delta_{ss'}, 
\quad 
 \bar u_{s'}(p')\gamma^+\gamma_5u_{s}(p) = 2\sqrt{p^+p'^+}\delta_{ss'} \mathrm{sign}(s).
\end{equation}

The spinor matrix elements for the one-gluon exchange are collected in Table~\ref{tab:spinor_matrix} (see also Table I of
Ref.~\cite{Wiecki:2014ola}).

\begin{table}
\centering 
\caption{Spinor matrix elements $\bar{u}_{s_1'}(p_1')\gamma_\mu u_{s_1}(p_1)\bar{v}_{s_2}(p_2)\gamma^\mu v_{s_2'}(p_2')$. 
$m_q$ ($m_a$) is the mass of the quark (antiquark). $x=p^+_1/P^+$ and $x'=p'^+_1/P^+$ are longitudinal momentum fractions of the quark, 
$\vec p_\perp = \vec p_{1\perp} - x\vec P_\perp$ and $\vec p'_\perp = \vec p'_{1\perp} - x'\vec P_\perp$ are relative transverse
momenta. For convenience, we use the complex representation for the transverse vectors, viz, $p \triangleq p_x + \imag p_y$ and $p^*
\triangleq p_x - \imag p_y$.}
\label{tab:spinor_matrix}
\begin{tabular}{cccc|c}
\toprule
\quad$s_1$\quad & \quad$s_2$\quad & \quad$s_1'$\quad & \quad$s_2'$\quad & $\mathlarger{\frac{\bar{u}_{s_1'}(p_1')\gamma_\mu
u_{s_1}(p_1)\bar{v}_{s_2}(p_2)\gamma^\mu
v_{s_2'}(p_2')}{2\sqrt{x(1-x)x'(1-x')}}} $ \\
\colrule
$+$&$+$&$+$&$+$& $m_q^2\frac{1}{xx'}+m^2_a\frac{1}{(1-x)(1-x')} + \frac{pp'^*}{x(1-x)x'(1-x')}$  \\
$-$&$-$&$-$&$-$&$ m_q^2\frac{1}{xx'}+m^2_a\frac{1}{(1-x)(1-x')} + \frac{p^*p'}{x(1-x)x'(1-x')}$ \\
\colrule
$+$&$-$&$+$&$-$&$
m_q^2\frac{1}{xx'}+m^2_a\frac{1}{(1-x)(1-x')} + \left(\frac{p'^\ast}{x'}+\frac{p^\ast}{1-x}\right)
\left(\frac{p}{x}+\frac{p'}{1-x'}\right)$  \\
$-$&$+$&$-$&$+$&$
m_q^2\frac{1}{xx'}+m^2_a\frac{1}{(1-x)(1-x')} + \left(\frac{p^\ast}{x}+\frac{p'^\ast}{1-x'}\right)
\left(\frac{p'}{x'}+\frac{p}{1-x}\right)$\\
\colrule
$+$&$+$&$+$&$-$&$m_a \frac{x'}{(1-x)(1-x')}\Big(\frac{p'}{x'} - \frac{p}{x} \Big)$\\
$-$&$-$&$-$&$+$&$m_a \frac{x'}{(1-x)(1-x')}\Big(\frac{p^*}{x} - \frac{p'^*}{x'} \Big)$\\
\colrule
$-$&$+$&$-$&$-$&$m_a \frac{x}{(1-x)(1-x')}\Big(\frac{p'}{x'} - \frac{p}{x} \Big)$\\
$+$&$-$&$+$&$+$&$m_a \frac{x}{(1-x)(1-x')}\Big(\frac{p^*}{x} - \frac{p'^*}{x'} \Big)$\\
\colrule
$+$&$+$&$-$&$+$&$m_q \frac{1-x'}{xx'}\Big(\frac{p}{1-x} - \frac{p'}{1-x'} \Big)$\\
$-$&$-$&$+$&$-$&$m_q \frac{1-x'}{xx'}\Big(\frac{p'^*}{1-x'} - \frac{p^*}{1-x} \Big)$\\
\colrule
$+$&$-$&$-$&$-$&$m_q \frac{1-x}{xx'}\Big(\frac{p}{1-x} - \frac{p'}{1-x'} \Big)$\\
$-$&$+$&$+$&$+$&$m_q \frac{1-x}{xx'}\Big(\frac{p'^*}{1-x'} - \frac{p^*}{1-x} \Big)$\\
\colrule
$+$&$-$&$-$&$+$& \multirow{2}{*}{$ -m_q m_a  \frac{(x-x')^2}{x(1-x)x'(1-x')} $}\\
$-$&$+$&$+$&$-$& \\
\colrule
$+$&$+$&$-$&$-$& \multirow{2}{*}{0}\\
$-$&$-$&$+$&$+$& \\
\botrule
\end{tabular}
\end{table}

\section{Polarization Vectors}

\paragraph{gauge bosons}
The polarization vector of a gauge boson in light-cone gauge $A^+ = 0$ is:
\begin{equation}
\varepsilon^\mu_\lambda(k) = (\varepsilon^-_\lambda, \varepsilon^+_\lambda, \vec \varepsilon_{\lambda\perp}) \triangleq 
\Big( \frac{2\vec\epsilon_{\lambda\perp} \cdot \vec k_\perp}{k^+}, 0, \vec \epsilon_{\lambda\perp} \Big) , \quad (\lambda = \pm 1)
\end{equation}
where $\vec\epsilon_{\pm\perp} = \frac{1}{\sqrt{2}}(-1, \mp \imag)$. The polarization vector defined here satisfies:
\begin{itemize}
 \item $k_\mu \varepsilon^\mu_\lambda(k) = 0$;
 \item $\varepsilon^\mu_\lambda(k)\varepsilon^*_{\lambda' \mu}(k) = - \delta_{\lambda,\lambda'}$;
 \item helicity sum:
\begin{equation}
\sum_{\lambda=\pm} \varepsilon^{\mu *}_\lambda(k)\varepsilon^\nu_\lambda(k) = -g^{\mu \nu} + \frac{n^\mu k^\nu + n^\nu
k^\mu}{n\cdot k} - \frac{k^2}{(n\cdot k)^2}n^\mu n^\nu.
\end{equation}
Here $n = (1, 0, 0, -1)$ is a light-like 4-vector ($n_\mu n^\mu=0$) perpendicular to the light front.
\end{itemize}

\paragraph{vector bosons}

The polarization vector for the a vector boson:
\begin{equation}
 e^\mu_\lambda(k) = \big(e^-_\lambda(k), e^+_\lambda(k), \vec e_{\lambda\perp}(k)\big) \triangleq 
\left\{ 
\begin{array}{lc}
\big(\frac{\vec k_\perp^2-m^2}{ m k^+}, \frac{k^+}{m}, \frac{\vec k_\perp}{m} \big), & \lambda = 0 \\
\big(\frac{2\vec \epsilon_{\lambda\perp} \cdot \vec k_\perp}{k^+}, 0, \vec \epsilon_{\lambda\perp} \big), & \lambda = \pm 1\\
\end{array}\right. 
\end{equation}
where $m$ is the mass of the vector boson. The polarization vector defined here satisfies:
\begin{itemize}
\item $k_\mu e_\lambda^\mu(k) = 0$;
 
\item $e^\mu_\lambda(k) e^*_{\lambda' \mu}(k) = - \delta_{\lambda,\lambda'}$;
\item spin sum:
\begin{equation}
\sum_{\lambda=0,\pm1} 	 e^{\mu *}_\lambda(k) e^\nu_\lambda(k) = -g^{\mu \nu} + \frac{k^\mu k^\nu}{k^2}.
\end{equation}

\end{itemize}

\end{document}